%% file: hvp_ccslo.tex
\documentclass[aps,prd,12pt,longbibliography,nofootinbib,tightenlines,preprintnumbers,superscriptaddress,notitlepage]{revtex4-1}
\usepackage{graphicx}
\graphicspath{{./Figures/}}

\usepackage[utf8]{inputenc}

\usepackage{amssymb}
\usepackage{amsmath}
\usepackage{xcolor}
\usepackage{hyperref}
\usepackage{multirow}
\usepackage{slashed}
\usepackage{subcaption}
\hypersetup{colorlinks, linkcolor = [rgb]{0,0.0,0.75}, citecolor = [rgb]{0,0.0,0.75}, urlcolor = [rgb]{0,0.0,0.75}}

\makeatletter
\g@addto@macro\bfseries{\boldmath}
\makeatother

\input{definitions}

\begin{document}

\preprint{MITP-22-084}

\title{Coordinate-space calculation of the window observable for the hadronic vacuum polarization contribution to $(g-2)_\mu$}

\author{En-Hung~Chao}
\affiliation{PRISMA$^+$ Cluster of Excellence \& Institut f\"ur Kernphysik,
Johannes Gutenberg-Universit\"at Mainz,
D-55099 Mainz, Germany}
\affiliation{Physics Department,
Columbia University,
New York, New York 10027, USA}

\author{Harvey~B.~Meyer}
\affiliation{PRISMA$^+$ Cluster of Excellence \& Institut f\"ur Kernphysik,
Johannes Gutenberg-Universit\"at Mainz,
D-55099 Mainz, Germany}
\affiliation{Helmholtz~Institut~Mainz,
Staudingerweg 18, D-55128 Mainz, Germany}
\affiliation{GSI Helmholtzzentrum f\"ur Schwerionenforschung, Darmstadt, Germany}

\author{Julian~Parrino}
\affiliation{PRISMA$^+$ Cluster of Excellence \& Institut f\"ur Kernphysik,
Johannes Gutenberg-Universit\"at Mainz,
D-55099 Mainz, Germany}

\begin{abstract}
  The `intermediate window quantity' of the hadronic vacuum polarization contribution to the anomalous magnetic moment of the muon
allows for a high-precision comparison between the data-driven approach and lattice QCD. The existing lattice results, which presently show good consistency among each other, are in strong tension with
  the data-driven determination. In order to check for a potentially common source of systematic error of the lattice calculations, which are all based on the time-momentum representation (TMR),  we perform a calculation using a Lorentz-covariant coordinate-space (CCS) representation.
We present results for the isovector and the connected strange-quark contributions to the intermediate window quantity at
a reference point in the $(m_\pi,m_K)$ plane, in the continuum and infinite-volume limit, based on four different lattice spacings.
Our results are in good agreement with those of the recent TMR-based Mainz-CLS publication.
\end{abstract}

\date{\today}

\maketitle

\input{intro}
\input{formalism}
\input{lattice_setup}

\input{fse_correction}

\input{result}
\input{conclu}

\acknowledgments{This work is supported by  the European Research Council (ERC) under the
  European Union's Horizon 2020 research and innovation programme
  through grant agreement 771971-SIMDAMA, as well as by  
 the Deutsche
  Forschungsgemeinschaft (DFG) through the Cluster of Excellence \emph{Precision Physics, Fundamental Interactions, and Structure of Matter} (PRISMA+ EXC 2118/1)  within the German Excellence Strategy (Project ID 39083149). 
 E.-H.C.'s work was supported in part by the U.S. D.O.E. grant \#DE-SC0011941.
Calculations for this project were partly performed on the HPC clusters ``Clover'' and ``HIMster II'' at the Helmholtz-Institut Mainz and ``Mogon II'' at JGU Mainz. 

Our programs use the deflated SAP+GCR solver from the openQCD package~\cite{Luscher:2012av}, as well as the QDP++ library
\cite{Edwards:2004sx}.
The measurement codes were developed based on the C++ library \texttt{wit}, an coding effort led by Renwick J.~Hudspith. 

We thank the authors of Ref. \cite{Ce:2022kxy} and especially Simon Kuberski for sharing the fit parameters  of the continuum extrapolation calculation in the TMR method, which we use in Sect. \ref{sect:shift}.
We are grateful to our colleagues in the CLS initiative for sharing ensembles.

\appendix
\input{window_kernel}

\input{fse_sakurai.tex}

\input{tables.tex}

\bibliography{refs}

\end{document}

%% file: definitions.tex
\renewcommand{\vec}[1]{\boldsymbol{#1}}
\newcommand{\be}{\begin{equation}}
\newcommand{\ee}{\end{equation}}
\newcommand{\ba}{\begin{eqnarray}}
\newcommand{\ea}{\end{eqnarray}}
\newcommand{\la}{\label}

\newcommand{\Tr}{\,\hbox{\rm Tr}}

\newcommand{\<}{\langle}
\renewcommand{\>}{\rangle}

\newcommand{\txts}{\textstyle}

\newcommand{\amu}{a_\mu}

\newcommand{\ahvp}{a_\mu^{\mathrm{hvp}} }
\newcommand{\alphaqed}{\alpha_{\mathrm{QED}}}
\newcommand{\aw}{a_\mu^{\textrm{W}}}

%% file: intro.tex
\section{Introduction}
As a precision observable, the anomalous magnetic moment of the muon, $a_\mu$, has  attracted a great deal of attention in recent years.
With the release of the first results by Fermilab's E989 experiment in 2021,
the experimental world-average~\cite{Muong-2:2006rrc, Muong-2:2021ojo} of this quantity has reached the precision level of 35\;ppm.
Tremendous efforts have also been invested on the theory side to reach the same level of precision.
To achieve the desired precision target, it is indispensable to bring the hadronic contributions
-- which entirely dominate the error budget of the theory estimate --  under good control.
The various hadronic contributions are classified according to the order in
the electromagnetic coupling constant $\alphaqed$ at which they contribute to $a_\mu$.
The leading, O$(\alphaqed^2)$ term is the hadronic vacuum polarization (HVP) contribution to $a_\mu$.
Together with the  O$(\alphaqed^3)$ 
 hadronic light-by-light contribution (HLbL), it has been the key quantity to improve over the last decade in order to reach
the precision that the direct experimental measurement would achieve in the near future.
The efforts from the theory community to resolve the hadronic contributions to $a_\mu$ can be sorted into two categories of methodology: the data-driven~\cite{davier:2017zfy, keshavarzi:2018mgv, colangelo:2018mtw, hoferichter:2019mqg, davier:2019can, keshavarzi:2019abf, kurz:2014wya, melnikov:2003xd, masjuan:2017tvw, Colangelo:2017fiz, hoferichter:2018kwz, bijnens:2019ghy, colangelo:2019uex, pauk:2014rta, danilkin:2016hnh, jegerlehner:2017gek,knecht:2018sci,eichmann:2019bqf,roig:2019reh,colangelo:2014qya} and lattice~\cite{chakraborty:2017tqp,borsanyi:2017zdw,blum:2018mom,giusti:2019xct,shintani:2019wai,FermilabLattice:2019ugu,gerardin:2019rua,Aubin:2019usy,giusti:2019hkz, Blum:2019ugy,gerardin:2019vio, Borsanyi:2020mff, Chao:2020kwq, Chao:2021tvp, Chao:2022xzg}.
The 2020 $(g-2)_\mu$ theory White Paper (WP)~\cite{Aoyama:2020ynm} provided the Standard Model prediction
at a precision level comparable to that of the experiment; that prediction currently stands in
4.2$\sigma$ tension with the experimental world-average for $a_\mu$.
To confirm the discrepancy, further improvement on the uncertainties is needed.
Especially, the HVP contribution $\ahvp$ has to be known to the few-per-mille level.

The WP value for the HVP is solely based on the data-driven method,
due to the lattice determinations having larger uncertainties at the
time of the publication. After the publication of the WP, the
Budapest-Marseille-Wuppertal (BMW) collaboration published their
lattice QCD estimate for $\ahvp$ at almost the same precision
level~\cite{Borsanyi:2020mff}.  Their calculation, however, yields a
larger value for $a_\mu$, in better agreement with the direct
experimental measurement.  Although their result should still be
verified by other lattice collaborations, preferably using different
discretization schemes to pin down potential systematic errors,
understanding the tension within SM predictions resulting from
different classes of methods has become a matter of high priority.
Especially the HVP contribution needs to be sharply scrutinized, as it
dominates currently the hadronic uncertainties.

The window quantities for $\ahvp$, originally introduced in
Ref.~\cite{RBC:2018dos}, provide a good way to break down the $\ahvp$
into subcontributions associated with different Euclidean time
intervals.  In particular, the \textit{intermediate window} suggested
therein is a more tractable observable for lattice practitioners, as
it avoids the short-distance region, where discretization effects can
become hard to control, and the large-distance region, where the
statistical Monte-Carlo noise and finite-size effects become the
limiting factor.  As the calculation of this observable is amenable
to the data-driven methods~\cite{Colangelo:2022vok}, the theory
community has invested significant effort into refining the estimates on this
quantity~\cite{RBC:2018dos,Borsanyi:2020mff,Wang:2022lkq,Aubin:2022hgm,Ce:2022kxy,Alexandrou:2022amy,FermilabLattice:2022smb}.
The original formulation of the window quantity in fact relies on the
Time-Momentum Representation (TMR) of $\ahvp$, which involves a
Euclidean-time correlation function calculated at vanishing spatial
momentum~\cite{Bernecker:2011gh}.  The aim of the present paper is to
offer a verification of the method based on an alternative formulation
which utilizes the position-space Euclidean-time two-point correlator
without any momentum projection.  This alternative makes use of the
previously introduced Covariant Coordinate-Space (CCS)
kernel~\cite{Meyer:2017hjv}, which is motivated by the rapid fall-off
of the Euclidean correlation function with the spacetime separation.
An important feature of this alternative formulation is the fact that
the lattice points are treated in an O$(4)$-covariant way, whereby
different discretization effects are expected than under the TMR.
Therefore, the CCS representation can provide a valuable check for the
continuum extrapolated value obtained from the TMR.  In this work, we
focus on lattice ensembles at an almost fixed pion mass of around
350\;MeV at four different lattice spacings and apply finite-size
corrections ensemble by ensemble based on a field-theoretic model
which is able to describe to a good degree the experimental data of
the pion electromagnetic form factor. On the one hand, this
calculation provides a proof-of-principle that the CCS method is not
only viable, but also quite competitive with the TMR method. On the
other hand, at $m_\pi=350\,$MeV we are able to directly compare our result to the recent
calculation by the Mainz-CLS collaboration~\cite{Ce:2022kxy}, thereby testing whether lattice-QCD based results
are independent of the chosen representation. Ultimately, this represents
a test of the restoration of Lorentz invariance, which is broken both
at short distances  by the lattice and in the infrared by the finite volume.


This paper is organized as follows.  In Section~\ref{sect:formalism},
we present the CCS formalism for the calculation of the window
quantities.  Our numerical setup and computational strategies are
reported in Section~\ref{sect:lattice_setup}.
Section~\ref{sect:fse_correction} is dedicated to the correction of
the finite-size effects used for this work.  The continuum
extrapolation of the results is discussed and compared to the previous
Mainz results~\cite{Ce:2022kxy} evaluated at the same pion mass in
Section~\ref{sect:result}.  Finally, concluding remarks are made in
Section~\ref{sect:conclu}.


%% file: formalism.tex
\section{Formalism}\label{sect:formalism}
Under the time-momentum representation (TMR)~\cite{Bernecker:2011gh}, the hadronic vacuum polarization contribution to $a_\mu$ can be written as an integral over the two-point function
\be
G_{\mu\nu}(x) = \langle j_\mu(x) j_\nu(0)\rangle
\ee
of the quark electromagnetic current
\be
j_\mu = \sum_f {\cal Q}_f \; \bar \psi_f \gamma_\mu \psi_f
\qquad \qquad ({\cal Q}_u={\txts\frac{2}{3}},\;{\cal Q}_d={\cal Q}_s={\txts-\frac{1}{3}}),
\ee
in Euclidean time weighted with a QED kernel~\cite{DellaMorte:2017dyu}.
Explicitly, the TMR representation of $a_\mu^{\rm hvp}$ reads
\be 
\label{a_mu_tmr}
a_\mu^{\rm hvp} = \Big(\frac{\alpha}{\pi}\Big)^2 \int_0^\infty dt \,f(t,m_\mu)\, {\cal G}(t)\,,
\ee 
where $G(t)$ is the two-point correlator projected to vanishing spatial momentum, 
\be 
\label{TMRcorrelator}
{\cal G}(t)\,\delta_{ij} = -  \int d^3x \;G_{ij}(t,\vec x)\,,
\ee 
and $f(t,m_\mu)$ is the QED kernel
\be 
f(t,m_\mu) = \frac{2\pi^2}{m_\mu^2}\Big(-2 +8\gamma_E +\frac{4}{\hat{t}^2} +\hat{t}^2 -\frac{8 K_1(2\hat{t})}{\hat{t}}+8\ln(\hat{t})+ G^{2,1}_{1,3}\left(\hat{t}^2 \Big| \begin{matrix} \frac{3}{2} \\ 0,1,\frac{1}{2}  \end{matrix} \right) \Big)\,,
\ee
where $\hat{t}\equiv tm_\mu$.

Although Eq.~\eqref{a_mu_tmr} provides a way to compute $\ahvp$ on the lattice, the necessity to precisely control the discretization effects stemming from small Euclidean-times and the loss of statistical quality in long Euclidean-times make it challenging for lattice calculations to achieve the same precision as methods using the $R$-ratio~\cite{Brodsky:1967sr, Lautrup:1968tdb}.
Alternative observables were first proposed in Ref.~\cite{RBC:2018dos}, which consist in filtering contributions from different Euclidean-time regions with appropriate extra weight factors to Eq.~\eqref{a_mu_tmr}.
One can alter the kernel with the help of smoothed Heaviside functions $\Theta_\Delta(t) = \frac{1}{2} (1+\tanh(\frac{t}{\Delta}))$ to restrict the integral to a particular energy window, which amounts to substituting the QED-kernel appearing in Eq.~\eqref{a_mu_tmr} by
\be\label{eq:winkernel}
f_{\rm{W}}(t,m_\mu) = \Big[ \Theta_\Delta(t-t_0) - \Theta_\Delta(t-t_1) \Big] f(t,m_\mu)\,.
\ee
The original proposal in Ref.~\cite{RBC:2018dos} was motivated by the fact that lattice calculations and phenomenological estimates can be made accurate in different euclidean time windows; applying different methods according to their performance in the concerned region can thus lead to a better combined estimate.
Typically, lattice calculations suffer from enhanced discretization effects at very short distances, and the long-distance nature of $\amu$ makes the finite-size corrections non-negligible. 
The \textit{intermediate window quantity}, $\aw$, defined by Eq.~\eqref{eq:winkernel} with $t_0 = 0.4$ fm, $t_1 =1.0$ fm and $\Delta = 0.15$ fm, is therefore expected to be well-suited for lattice calculations, where a sub-percent precision level with well-controlled systematic errors is easier to achieve than for the whole $\ahvp$.
A comparison between the lattice and phenomenological determinations of this quantity would shed light on the current tension within SM predictions since the publication of the BMW result~\cite{Colangelo:2022vok}.

During the past few years, many lattice results on $\aw$ have been published by independent collaborations~\cite{Aubin:2019usy, Borsanyi:2020mff, Lehner:2020crt,Lahert:2021xxu, Ce:2022kxy, Alexandrou:2022amy, Wang:2022lkq, Aubin:2022hgm}.
The current estimates from different lattice discretization schemes show nice consistency within the reached accuracy.
It is then worth checking the current available results, all obtained from the TMR, with alternative approaches to exclude a potential common bias from the TMR method.
Specifically, it is interesting to see if one can still get a consistent result from a method which has different discretization effects than the TMR. 
We propose an alternative representation of $\aw$ based on an alternative approach for the calculation of $a_\mu$, the Covariant Coordinate-Space (CCS) method, introduced in Ref.~\cite{Meyer:2017hjv}.
The derivation is given in Appendix~\ref{appendix:window_kernel}.
Qualitatively, it follows closely the derivation of the original CCS kernels, which applies to observables which can be written as a weighted integral over the Adler function $\mathcal{A}(Q^2) \equiv Q^2\frac{d}{dQ^2}\Pi(Q^2)$, where $\Pi(Q^2)$ is the vacuum polarization function.
Non-trivial examples of such observables are the subtracted Vacuum Polarization function and $\amu$.

Exploiting the transversality of the vacuum polarization tensor, the dependence on the tensor structure of the vector-vector correlator $G_{\mu\nu}(x) $ can be made explicit and we arrive at a representation of $\aw$ as a four-dimensional integral,
\be
\label{CCSIntegral}
a^\textrm{W}_\mu =  \int d^4x\; H_{\mu\nu}(x)\, G_{\mu\nu}(x)\,,
\ee 

where the symmetric, rank-two, transverse ($\partial_\mu H_{\mu\nu}=0$) kernel
\be
\label{eq:ccs-kernel}
H_{\mu \nu}(x) = -\delta_{\mu\nu}\mathcal{H}_1(|x|) +\frac{x_\mu x_\nu}{|x|^2}\mathcal{H}_2(|x|)
\ee
is characterized by two scalar weight functions,
\be
\mathcal{H}_1(|x|) = \frac{2}{9\pi|x|^4}\int_0^{|x|}dt \sqrt{|x|^2-t^2}(2|x|^2+t^2)f_{\rm{W}}(t,m_\mu)\,,
\ee 
\be 
\mathcal{H}_2(|x|) = \frac{2}{9\pi|x|^4}\int_0^{|x|}dt \sqrt{|x|^2-t^2}(4t^2-|x|^2)f_{\rm{W}}(t,m_\mu)\,.
\ee

A remarkable feature of the CCS method is the possibility of modifying the kernel and hence the integrand in Eq.~\eqref{CCSIntegral} without changing the final integrated value in infinite-volume, thanks to current conservation. 
Effectively, using the fact that the vector-vector correlator is conserved, $\partial_\mu G_{\mu \nu}=0$, one can add a total-derivative term of type $\partial_\mu [x_\nu g(|x|)]$ to the kernel without changing $\aw$, as this only leads to a surface term vanishing in infinite volume~\cite{Ce:2018ziv}.

This flexibility makes lattice calculations with the CCS method attractive because it allows one to find an optimum in terms of discretization- and finite-size-errors by controlling the sensitivity of the integrand to different regions by adjusting the shape of the integrand (see Sect.~\ref{sect:lattice_setup} for our setup for the lattice computation).
In particular, the success in the control of the finite-size effects in the Hadronic Light-by-Light contribution to $a_\mu$ in an analogous way~\cite{Blum:2016lnc, Asmussen:2019act} makes this a promising strategy.
Nonetheless, the systematic error induced by finite-size and discretization effects might require careful studies for each kernel. These are important subjects of this paper (Sect.~\ref{sect:fse_correction} and Sect.~\ref{sect:result}).

In the following, we will perform calculations with two additional kernels, the traceless 
\be
\label{TracelessKernel}
H^{\textrm{TL}}_{\mu\nu}(x) = \left(-\delta_{\mu \nu} +4\frac{x_\mu x_\nu}{|x|^2}\right)\mathcal{H}_2(|x|)\,,
\qquad {\textrm{ (`TL')}}
\ee 
and the one which is proportional to $x_\mu x_\nu$,
\be 
\label{XXKernel}
H^{\textrm{XX}}_{\mu\nu}(x) = \frac{x_\mu x_\nu}{|x|^2}\Big(\mathcal{H}_2(|x|) +|x|\frac{d}{d|x|}\mathcal{H}_1(|x|) \Big)
\qquad {\textrm{ (`XX')}}.
\ee
These choices were studied in Ref.~\cite{Ce:2018ziv}.
In particular, the XX kernel is motivated by a stronger suppression of the contributions from long distances when the correlator is modeled by a simple vector-meson exchange~\cite{Meyer:2017hjv}.
Finally, for the remainder of the paper, we denote a generic kernel as
\be 
\widetilde{H}_{\mu \nu}(x) = -\delta_{\mu\nu}\widetilde{\mathcal{H}}_1(|x|) +\frac{x_\mu x_\nu}{|x|^2}\widetilde{\mathcal{H}}_2(|x|)\,.
\ee 

%% file: lattice_setup.tex
\section{Lattice setup}\label{sect:lattice_setup}
We apply the CCS method to five different $N_f=2+1$ flavor gauge ensembles generated by the Coordinated Lattice Simulations consortium~\cite{Bruno:2014jqa} at a pion mass around $350$ MeV.
These ensembles have been generated with the O$(a)$-improved Wilson-clover fermion action and tree-level O$(a^2)$ improved L\"uscher-Weisz gauge action.
The detailed information about the used ensembles can be found in Tab.~\ref{table:ensemble}.
In this work, our goal is to provide a cross-check for the calculation carried out in the conventional TMR method~\cite{Ce:2022kxy}, restricting ourselves to the (strongly dominant) quark-connected contributions in the $f=u,d,s$ sector.

To control the discretization effects, in this work, we consider both the local (L) and the conserved (C) version
of the vector current on the lattice
\begin{equation} 
  j_\mu^{\textrm{(L)}}(x) =  
  \bar{\psi}(x)\gamma_\mu {\cal Q} {\psi}(x)\,,
\end{equation}
and 
\begin{eqnarray}
    j_\mu^{\textrm{(C)}}(x) &=& {\textstyle\frac{1}{2}} \left(j_\mu^{\textrm{(N)}}(x) + j_\mu^{\textrm{(N)}}(x-a\hat\mu) \right),
  \\
  j_\mu^{\textrm{(N)}}(x) &=& \frac{1}{2}\Big[\bar{\psi}(x+a\hat{\mu}) (1+\gamma_\mu)U_\mu^\dagger(x) {\cal Q}{\psi}(x)
	 -\bar{\psi}(x)(1-\gamma_\mu)U_\mu(x){\cal Q}{\psi}(x+a\hat{\mu})\Big]\,,
\end{eqnarray}
where $U_\mu(x)$ is the gauge link and ${\cal Q}$ is a generic quark charge matrix acting in flavor space. Starting from the Noether current $j_\mu^{\textrm{(N)}}$, we have defined the site-centered current $j_\mu^{\textrm{(C)}}$, which obeys the on-shell conservation equation $\sum_{\mu=0}^3 \partial_\mu^* \,j_\mu^{\textrm{(C)}} = 0$, where $\partial_\mu^*$ is the lattice backward derivative.

In practice, to handle the O$(a)$ lattice artifacts, we substitute the lattice vector currents with their improved counterparts\footnote{Eq.\ (\ref{eq:impr}) is valid for the flavor non-singlet combinations considered here.}~\cite{Bhattacharya:2005rb}
\be\label{eq:impr}
j^{(\alpha),\textrm{I}}_\mu(x) = j^{(\alpha)}_\mu(x) +ac^{(\alpha)}_V \partial_\nu T_{\mu \nu}(x),\qquad \textrm{for}\quad \alpha = \textrm{L, C}\,,
\ee
where the local tensor current is defined by $T_{\mu \nu}\equiv-\frac{1}{2}\bar{\psi}(x)[\gamma_\mu, \gamma_\nu]{\cal Q}\psi(x)$ and $c^{(\alpha)}_V$ is an improvement coefficient. 
For $c^{(\alpha)}_{V}$, we use the interpolating formulae Eq.~(46.a) and Eq.~(46.b) of Ref.~\cite{Gerardin:2018kpy},
consistently with the treatment of Ref.~\cite{Ce:2022kxy}.
For both flavor combinations considered here, the renormalization is multiplicative,
\ba\label{eq:renorm}
j^{({\rm L}),\textrm{R}}_\mu(x) = \hat{Z}_{\rm V}^{\textrm{(L)}}\, j^{({\rm L}),\textrm{I}}_\mu(x).
\ea

In the case of the local isovector current, corresponding to ${\cal Q} = {\rm diag}(\frac{1}{2},-\frac{1}{2},0)$,
the renormalization factor is given by
\ba\label{eq:renormI1}
\hat{Z}_{\rm V}^{\textrm{(L)}} = Z_V(g_0)\Big[1+3\bar{b}_V^{\rm eff} a m_{\rm q}^{\textrm{av}}+b_Vam_{{\rm q},l}\Big]\,,
\ea 
where the parameters $Z_V$, $\bar{b}_V^{\rm eff}$ and $b_V$ are obtained from the Pad\'e fits Eqs.~(44.a,b,c) of Ref.~\cite{Gerardin:2018kpy}. The average quark mass $m_{\rm q}^{\textrm{av}}$ and the mass of the quark of flavour $f$, $m_{{\rm q},f}$, are taken from the same reference. The conserved vector current does not need to be renormalized, thus we have $\hat{Z}_{\rm V}^{\textrm{(C)}}=1$.
This treatment of the renormalization and improvement coefficients corresponds to Set~1
in the recent calculation of the window observable of the Mainz group \cite{Ce:2022kxy}.

The strange current, since we consider only the connected contribution
to its two-point function, must be defined within a partially quenched
theory. For instance, adding a fourth, purely `valence' quark $s'$ mass-degenerate with $s$,
the flavor structure corresponds to ${\cal Q}={\rm diag}(0,0,\frac{1}{3\sqrt{2}},-\frac{1}{3\sqrt{2}})$.
The corresponding renormalization factor can be written in the form
\ba
\hat{Z}_{\rm V}^{\textrm{(L)}} = Z_V(g_0)\Big[1+3\bar{b}_V^{\rm eff} a m_{\rm q}^{\textrm{av}}
   + b_V am_{{\rm q},s}  + b^{\rm pq}_V (a m_{\rm q}^{\textrm{av}} - am_{{\rm  q},s})\Big]\,.
\ea
It contains an additional term  with a
coefficient $b^{\rm pq}_V$ (of order $g_0^4$ in perturbation theory) representing a
sea-quark effect.  Both for the latter reason and the fact that we work
quite close to the $SU(3)_{\rm f}$ point $a m_{\rm q}^{\textrm{av}} =
am_{{\rm q},s}$, we neglect this additional term.

We give some further details for the implementation in the following subsections.
Although the expressions are given for the case where both currents are local, the generalization to the cases with conserved currents should be straightforward.

\begin{table}
\begin{tabular}{|c|c|c|c|c|c|c|c|c|}
\toprule
Id   & $\beta$ & $L^3 \times T$ & $a$ [fm] & $m_\pi$ [MeV] & $m_K$ [MeV] & $m_\pi L$ & $L$ [fm] & \begin{tabular}[c]{@{}l@{}}\#confs \\ light/strange\end{tabular} \\ \hline
U102 & 3.4     & $24^3 \times 96$              & 0.08636    & 353(4)            & 438(4)          & 3.7       & 2.1          & 200/0                                                               \\
H102 &         & $32^3 \times 96$              &            &                   &                 & 4.9       & 2.8          & 240/120                                                          \\ \hline
S400 & 3.46    & $32^3 \times 128$             & 0.07634    & 350(4)            & 440(4)          & 4.2       & 2.4          & 240/120                                                          \\ \hline
N203 & 3.55    & $48^3 \times 128$             & 0.06426    & 346(4)            & 442(5)          & 5.4       & 3.1          & $90\times 2$/$90\times 2$                                                             \\ \hline
N302 & 3.7     & $48^3 \times 128$             & 0.04981    & 346(4)            & 450(5)          & 4.2       & 2.4          & 240/120                                                          \\ 
\botrule
\end{tabular}
	\caption{Overview of the used ensembles. The lattice spacings are determined in Ref.~\cite{Bruno:2016plf} and the pion and kaon masses are taken from Ref.~\cite{Ce:2022kxy}. Open boundary conditions are employed for all of the listed ensembles. For the ensemble N203, two replica have been included in the analysis. To exploit translational invariance to reduce statistical fluctuations, all contracted correlators [Eqs.~(\ref{eq:g1lat},\ref{eq:g2lat},\ref{eq:g3lat})] have been computed at $L$ different choices of origin situated at $(n,n,n,T/2)$.}

\label{table:ensemble}
\end{table}

\subsection{Contracted Correlators}
\label{sect:con_cor}
From the Lorentz structure of the CCS kernel Eq.~\eqref{eq:ccs-kernel}, we deduce that the integral representation of $\aw$, Eq.~\eqref{CCSIntegral}, can be conveniently written as 
\begin{equation}\label{eq:awlcc}
	\aw = \int_0^\infty dr\, f(r)\,,\quad
	f(r) \equiv \,r^3\Big[ - \widetilde{\mathcal{H}}_1(r) G_1(r) + \frac{1}{r^2} \widetilde{\mathcal{H}}_2(r) G_2(r) \Big]\,,
\end{equation}
where 
\be\label{eq:awlccg1} 
G_1(r) = \int_{\mathbb{S}^3} d\Omega_{x}\,  G_{\mu\nu}(x) \delta_{\mu \nu}\,,
\ee 
\be\label{eq:awlccg2} 
G_2(r) = \int_{\mathbb{S}^3} d\Omega_{x}\,  G_{\mu\nu}(x) x_\mu x_\nu\,,
\ee 
with $r\equiv |x|$, $\hat{x}\equiv x/|x|$ and $\mathbb{S}^3$ is the measure of the three-sphere.
The functions $G_1$ and $G_2$ will be referred to as the \textit{contracted correlators} and $f$ as the \textit{integrand}.

In infinite volume, the integrand transforms as a scalar under O$(4)$-transformations. 
In particular, it is expected to decay exponentially with the separation $r$ at large distances due to the behavior of the vector-current two-point correlator. 
For our lattice calculation, where the O$(4)$-symmetry is broken, the contracted correlators need to be sampled by points which are spread around on the same shell as evenly as possible to restore the rotational symmetry.
This in part motivates our choice for saving the following quantities on each given distance $r$ on the lattice for the quark-connected contribution of $\aw$
\be
\label{eq:g1lat}
\widehat{G}^{\textrm{conn.}}_1(r) = -\,{\rm Tr}\{{\cal Q}^2\} \sum_{ x\in\Lambda,\,|x| = r } \Re \Tr[ S(x,0) \gamma_\mu S(0,x) \gamma_\mu ]\,,
\ee 
\be 
\label{eq:g2lat}
\widehat{G}^{\textrm{conn.}}_2(r) = -\,{\rm Tr}\{{\cal Q}^2\} \sum_{ x\in\Lambda,\,|x| = r } \Re \Tr[ S(x,0) \slashed{x} S(0,x) \slashed{x} ]\,,
\ee
where $\Lambda$ denotes the set of all points on the lattice and $S(x,0)$ is a quark propagator with point-source at $0$.
Note that, in this convention, we have $\widehat{G}^{\textrm{conn.}}_i(r)\rightarrow r^3G_i(r)$ in the continuum and infinite-volume limit.
Another advantage of such choice is the re-usability of the data for other quantities for which the form factors of the CCS kernel are known; 
it suffices to substitute the form factors $\widetilde{\mathcal{H}}_1$ and $\widetilde{\mathcal{H}}_2$ in the master formula Eq.~\eqref{eq:awlcc} with the desired one in such a case.

For the $O(a)$-improvement of the discretized lattice vector current Eq.~\eqref{eq:impr}, there is another quantity which has to be taken into account due to the tensor current.
Starting with the O$(a)$-improved vector-current given Eq.~\eqref{eq:impr}, one can keep the explicit coefficient $ac_V$ fixed and substitute the vector- and tensor-currents by their continuum and infinite-volume limit counterparts.
Plugging it into the original infinite-volume vector-current two-point correlator, Eq.~\eqref{CCSIntegral} is then modified to, up to O$(a^2)$-terms, 
\be
\tilde{a}^\textrm{W}_\mu(a) =  \int d^4x\,\Big\{
H_{\mu\nu}(x) G_{\mu\nu}(x)+ac_V\Big[\langle j_\mu(x)T_{\nu \alpha}(0)\rangle-\langle T_{\mu \alpha}(x)j_\nu(0)\rangle\Big] \partial_\alpha H_{\mu\nu}(x)
\Big\}\,,
\ee
where we have performed an integration-by-part to get the second term on the right-hand side.
The second term in the curly bracket can be seen as a lattice artifact as it vanishes at the $a\rightarrow 0 $ limit at fixed $c_V$, where $\aw$ is recovered.
Exploiting the Lorentz symmetry as done previously, we can consider it as a convolution of the correlation function in the square-bracket as
\be
ac_V \int_0^{\infty}dr\,r\,\widetilde{\mathcal{H}}_3(r) G_3(r)\,, 
\ee
where 
\be
\label{eq:g3lat}
G_3(r) = \int_{\mathbb{S}^3}d\Omega_{x}\, x_\alpha\; \Big[-\langle j_\mu(x)T_{\mu \alpha}(0)\rangle+\langle T_{\mu\alpha}(x)j_\mu(0)\rangle\Big] ,
\ee
\be
\widetilde{\mathcal{H}}_3(r) = \widetilde{\mathcal{H}}_2(r) +r\widetilde{\mathcal{H}}^\prime_1(r)\,.
\ee

This observation facilitates the numerical computation as the same propagators required for the calculation of the previously-mentioned contracted correlators can be reused and leads to the quantity to be computed on the lattice
\be 
\widehat{G}_3^{\textrm{conn.}}(r^2) = -\,{\rm Tr}\{{\cal Q}^2\} \sum_{x\in\Lambda,\, |x| = r } \Re \Tr[ S(x,0) \gamma_\mu S(0,x) (\slashed{x}\gamma_\mu -  \gamma_\mu \slashed{x} ) ]\,.
\ee

\subsection{Summation Schemes}\label{sect:summation}

\begin{figure}[]
    \centering
    \includegraphics[width=0.7\textwidth]{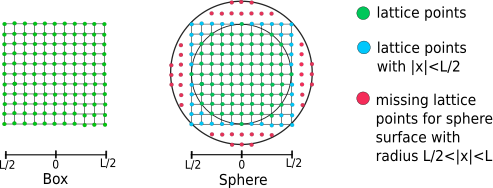}
    \caption{Visualization of the domain of integration on a hypercube of size $L$. The Details of the integration procedure are provided in Sect. \ref{sect:summation}}
    \label{fig:integration_scheme}
\end{figure}

Because of the periodicity in the spatial directions on the lattice, the spatial separation in each direction is mapped to
$x_k \in [-L/2,\, L/2]$ in infinite-volume spacetime.
This means that, in total, one can sample up to $r=L$ on a lattice with $T\geq L$.
However, the CCS formulation consists in treating the lattice points shell-by-shell with fixed $r$ across the hypercube, following the radial direction;
in the $r>L/2$ region, the corresponding shell on the hypercube is not faithfully sampled anymore.
Upon taking the continuum and infinite-volume limit, the summations in the lattice-summed contracted correlators $\widehat{G}_i$ run over the three-sphere $\mathbb{S}^3$.
As the summands become O$(4)$-invariant objects in this limit, it suffices to evaluate them at a given point and multiply by the $\mathbb{S}^3$-measure to get the answer.
However, on a finite lattice, this simplified procedure is exposed to both discretization and finite-volume effects.
This is better illustrated with Fig.~\ref{fig:integration_scheme}: when going beyond $r=L/2$ in the radial direction, the hypersphere only intersects with a subset of points on the entire shell of the hypercube submerged in infinite-volume spacetime.

In order to control the finite-volume effects, we propose the following summation scheme for our lattice data. 
We correct for these missing points by a multiplicative factor given by:
\be 
\label{eq:r4}
c(r,L) = \frac{r_4((r/a)^2)}{n_{\textrm{avail}}(r^2,L)}\,, \quad \textrm{with} \quad r_4(n)=8\sum_{d\,\mid\,n,\ 4\,\nmid\,d}d\,
\ee 
being the number of ways to represent $n$ as the sum of four squares and $n_{\textrm{avail}}(r^2,L)$ is the number of available points on the lattice, which can easily be counted. 
The sum in Eq.~\eqref{eq:r4} runs over all divisors $d$ of the integer number $n$, where 4 is not a divisor of $d$ itself. This is known as Jacobi's four-square theorem. 
A proof is given for example in Ref.~\cite{doi:10.1080/00029890.2000.12005191}.
Note that, in this definition, $c(r,L)=1$ for all $r\leq L/2$.
This summation scheme allows one to sample the contribution from the portion of a hypersphere cut out by the box as described by the red points on the right panel of Fig.~\ref{fig:integration_scheme}.
As a consequence, our lattice version of the master formula for $\aw$ reads:
\be\label{eq:awintegrand}
\aw{}^{\textrm{,lat.}} = a^4\sum_{r=0}^{L}c(r,L) f^{\textrm{lat.}}(r)\,,
\ee
where the lattice integrand is defined as
\be
f^{\textrm{lat.}}(r) \equiv 
-\widetilde{\mathcal{H}}_1(r)\widehat{G}^{\textrm{conn.}}_1(r) + \frac{1}{r^2}\widetilde{\mathcal{H}}_2(r)\widehat{G}^{\textrm{conn.}}_2(r) + \frac{ac_V}{r^2}\widetilde{\mathcal{H}}_3(r)\widehat{G}^{\textrm{conn.}}_3(r)
\,.
\ee
The results for the ensembles given in Tab. \ref{table:ensemble} calculated with this scheme are collected in Tab.~\ref{table:results_ensemble}.
Finally, as commented early, we could also have computed the lattice-summed correlators $\widehat{G}_i$'s by
starting from the continuum expression (\ref{eq:awlcc}), which is based on 4d spherical coordinates, and implementing it in one particular direction on the lattice.
The result should agree with the summation scheme of Eq.\ (\ref{eq:awintegrand}) after a proper continuum and infinite-volume extrapolation.
In general, the two approaches introduce a different scaling toward to continuum limit.
We have explicitly verified in the present case that the difference between these two treatments
of the lattice data is much smaller than the statistical error of the data.

%% file: fse_correction.tex
\section{Correction for the finite-size effects}\label{sect:fse_correction}
The finite-size effects (FSEs) on the electromagnetic correlator 
come dominantly from the two-pion intermediate states, which belong to the isovector channel.
In the context of the TMR method, a number of different approaches have been considered to estimate the FSEs.
Perhaps the most straightforward way to estimate FSEs is to rely on Chiral Perturbation Theory (ChPT) in a finite box.
The role of the $\rho$-meson, which contributes very strongly to the HVP at intermediate distances, however only enters at higher orders~\cite{Aubin:2019usy}.
Alternatively, one can use phenomenological models, e.g. Ref.~\cite{Jegerlehner:2011ti}, to include the effects of the $\rho$~\cite{Borsanyi:2020mff}.
Finite-size effects in the tail of the TMR correlator can also be computed based on the pion electric form factor in the timelike region,
which can be obtained from auxiliary lattice calculations~\cite{Meyer:2011um,Francis:2013fzp,DellaMorte:2017dyu,gerardin:2019rua}.
Finally, the first terms of a systematic asymptotic expansion are given in Refs.~\cite{Hansen:2019rbh, Hansen:2020whp}, where the FSEs correction to $\ahvp$ are related to a pion-photon Compton scattering amplitude.

In our approach with the CCS method, where the position-space vector-vector correlator is needed, a new aspect in the study of volume effects comes from the Lorentz structure of the correlator as a symmetric rank-2 tensor under the breaking of the O(4)-symmetry into that of a subgroup of the hypercubic group H(4), or the octahedral group $O_h$ if the time extent is taken to be infinite.
In addition, it is not straightforward to generalize the approach of Refs.~\cite{Hansen:2019rbh, Hansen:2020whp} or of Ref.~\cite{Meyer:2011um}:
as the correlator used in the CCS method is a position-space object, the whole range of center-of-mass momenta must be considered.
For these reasons, we opted to base our FSEs estimate on the model proposed in Ref.~\cite{Jegerlehner:2011ti}.
We will refer to this model as the \textit{Sakurai QFT} in the remainder of the paper.

The pion electric form factor, $F_\pi$, is commonly parametrized by the Gounaris-Sakurai (GS) formula~\cite{Gounaris:1968mw}.
In particular, it incorporates different dominant vector resonances with their widths into the form factor.
Ref.~\cite{Jegerlehner:2011ti} suggests a model which is realistic at $\sqrt{s}<1$\;GeV:
the Lagrangian of the theory in Euclidean spacetime is given by
\be\la{eq:LEinit}
{\cal L}_E = \frac{1}{4}F_{\mu\nu}(A)^2 + \frac{1}{4} F_{\mu\nu}(\rho)^2 + \frac{1}{2} m_\rho^2 \rho_\mu^2
 + \frac{e}{2g_\gamma} F_{\mu\nu}(A)F_{\mu\nu}(\rho) + (D_\mu\pi)^\dagger (D_\mu\pi) + m_\pi^2 \pi^\dagger \pi,
\ee
with the covariant derivative $D_\mu  \equiv \partial_\mu -ieA_\mu -ig\rho_\mu$.
The degrees of freedom are the photon $A_\mu$, the pion $\pi$ and the massive $\rho$-meson $\rho_\mu$.
In this Lagrangian, the $\rho$-meson and the photon mix already at treelevel via the product of the field strengths,
known as kinetic mixing term.
The normalization condition $F_\pi(0)=0$ emerges as a result of gauge invariance, independently of the values of
the coupling constants $g_\gamma$ and $g$.
As a condition to determine the latter, we match the decay rates of the vector meson 
to $\pi^+\pi^-$ and to $e^+e^-$ to their experimentally measured values.
This procedure gives $g=5.98$ and $g_\gamma =4.97$ [Eq.~\eqref{eq:sakurig} and Eq.~\eqref{eq:sakuriggamma}].
The details of this derivation as well as the renormalization of the theory in infinite volume are deferred to Appendix~\ref{appendix:fse_sakurai}.

As a sanity check, we have looked at the predictions of the Sakurai QFT for different Euclidean time windows, i.e., different choices of $t_0$ and $t_1$ in Eq.~\eqref{eq:winkernel} at fixed $\Delta = 0.15$ fm.
With our choice of parameters $g$ and $g_\gamma$, the two-pion channel contribution to these windows computed in the Sakurai QFT to one-loop agrees surprisingly well with the analysis based on the $e^+e^-$ cross-section data below 1 GeV~\cite{Colangelo:2022vok}; see Tab.~\ref{table:Sakurai_time_windows}.
Note that according to the analysis presented in Ref.~\cite{Colangelo:2022vok}, the two-pion channel amounts about 70\% of the total $\ahvp$.
This observation further strengthens our confidence in the model.

\begin{table}[]
\begin{tabular}{|c|c|c|}
\toprule
	$[t_0, t_1]$ & Sakurai QFT & Ref.~\cite{Colangelo:2022vok} \\ \hline
	$[0,0.1]$ fm      & 0.66     & 0.83(1) \\ \hline
	$[0.1,0.4]$ fm    & 14.05    & 12.89(12) \\ \hline
	$[0.4,0.7]$ fm    & 53.03   & 51.02(45) \\ \hline
	$[0.7,1.0]$ fm    & 87.59   & 87.28(72) \\ \hline
	$[1.0,1.3]$ fm    & 94.05    & 95.31(73) \\ \hline
	$[1.3,1.6]$ fm    & 79.64    & 80.88(58) \\ \hline
	$[1.6,\infty]$ fm & 165.81   & 166.08(106) \\ \hline \hline
	total            & 494.83   &  494.30(355)\\
\botrule
\end{tabular}
	\caption{Predictions of the Sakurai QFT for different Euclidean time windows defined by Eq.~\eqref{eq:winkernel} with $\Delta=0.15$ fm and the corresponding values for $t_1$ and $t_0$. $m_\pi$ and $m_\rho$ in the Lagrangian are set to their physical values and $(g,g_\gamma)=(5.984,4.97)$. The precision requirement for the numerical integration is set below the displayed digits. All numbers in the table are in units of $10^{-10}$. The uncertainties quoted for the values from Ref.~\cite{Colangelo:2022vok} result from all sources of error added in quadrature.}
\label{table:Sakurai_time_windows}
\end{table}

\begin{figure}
    \centering
    \includegraphics[width=0.7\textwidth]{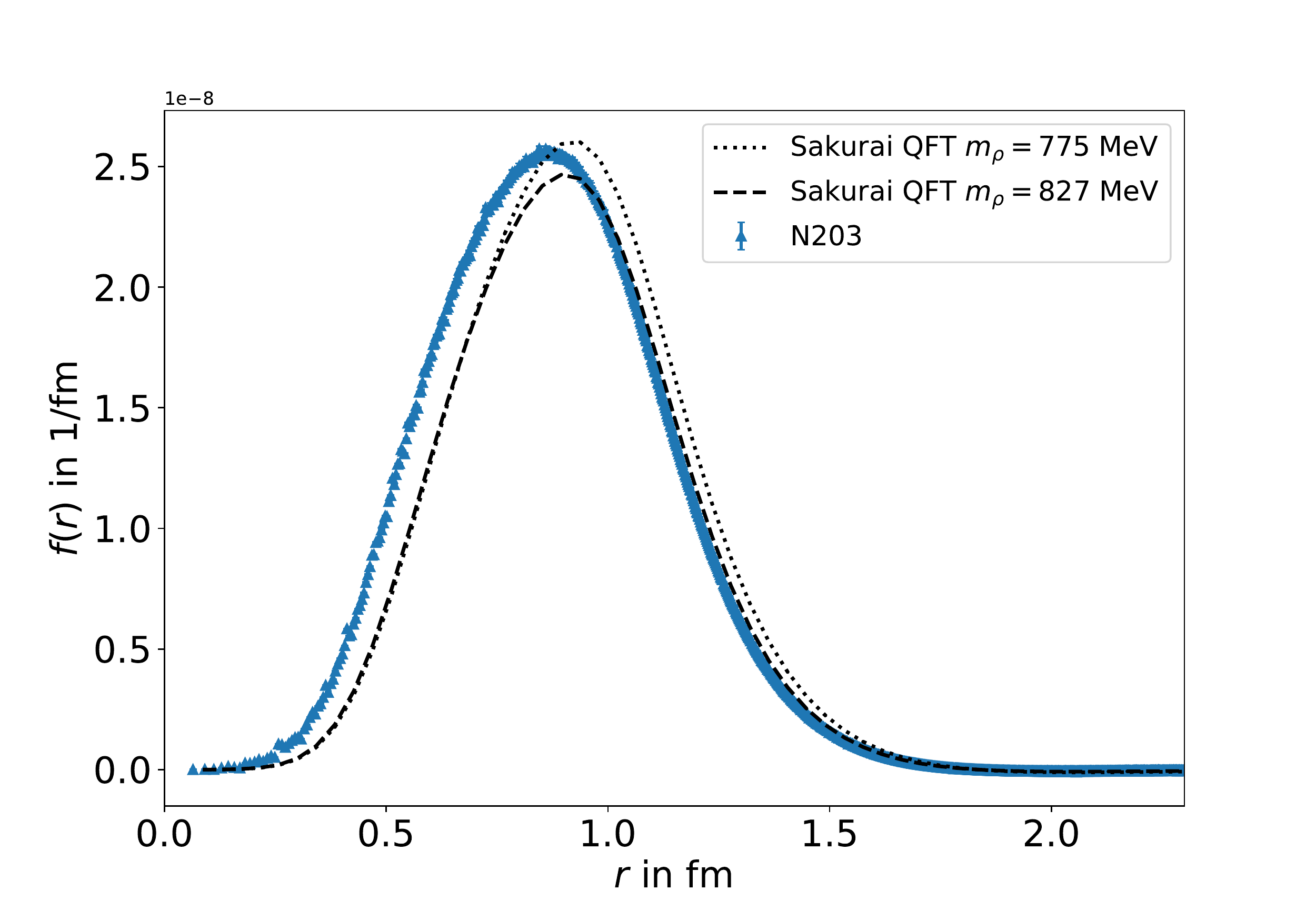}
	\caption{Comparison between the integrand from the ensemble N203 for the conserved-local discretization of the vector current and the prediction of the Sakurai QFT for the corresponding $m_\pi$ and $m_\rho$. The correction for the wrap-around-the-world pion has been applied to the lattice data.}
    \label{fig:sakurai_integrand}
\end{figure}

In Fig.~\ref{fig:sakurai_integrand}, we plot the infinite-volume integrands defined in Eq.~\eqref{eq:awlcc} predicted by the Sakurai QFT together with the finite-size corrected lattice integrand Eq.~\eqref{eq:awintegrand} for the ensemble N203, according to the procedure described in Sect.~\ref{sect:fsescheme}.
Two different $m_\rho$ are considered: one corresponds to its physical value (775 MeV) and the other (827 MeV) is obtained from a previous lattice study of the pion electric form factor in Gounaris-Sakurai parametrization~\cite{gerardin:2019rua}, evaluated at $m_\pi=350$ MeV.
A point worth mentioning  is the sensitivity to $m_\rho$.
At the considered pion mass, $m_\rho=827$ MeV gives an $\aw$ of $\sim 155\times 10^{-10}$, which is about 6\% lower than the value from $m_\rho=775$ MeV. 
This difference results from the different height of the peaks of the integrands. 
More importantly, as can be seen in the shape of the integrand in Fig.~\ref{fig:sakurai_integrand}, tuning the $\rho$-mass to its exact value predicted by the lattice study of Ref.~\cite{gerardin:2019rua} leads to a much better agreement in the long-distance region with the lattice data obtained in our study. 
In our study of the finite-size effects presented in this work, we have chosen $m_\rho$ to match the values listed in Ref.~\cite{gerardin:2019rua}. 

\subsection{Finite-size-effect correction scheme}
\label{sect:fsescheme}
\begin{figure}
\begin{subfigure}{0.49\textwidth}
     \centering
    \includegraphics[width=0.9\textwidth]{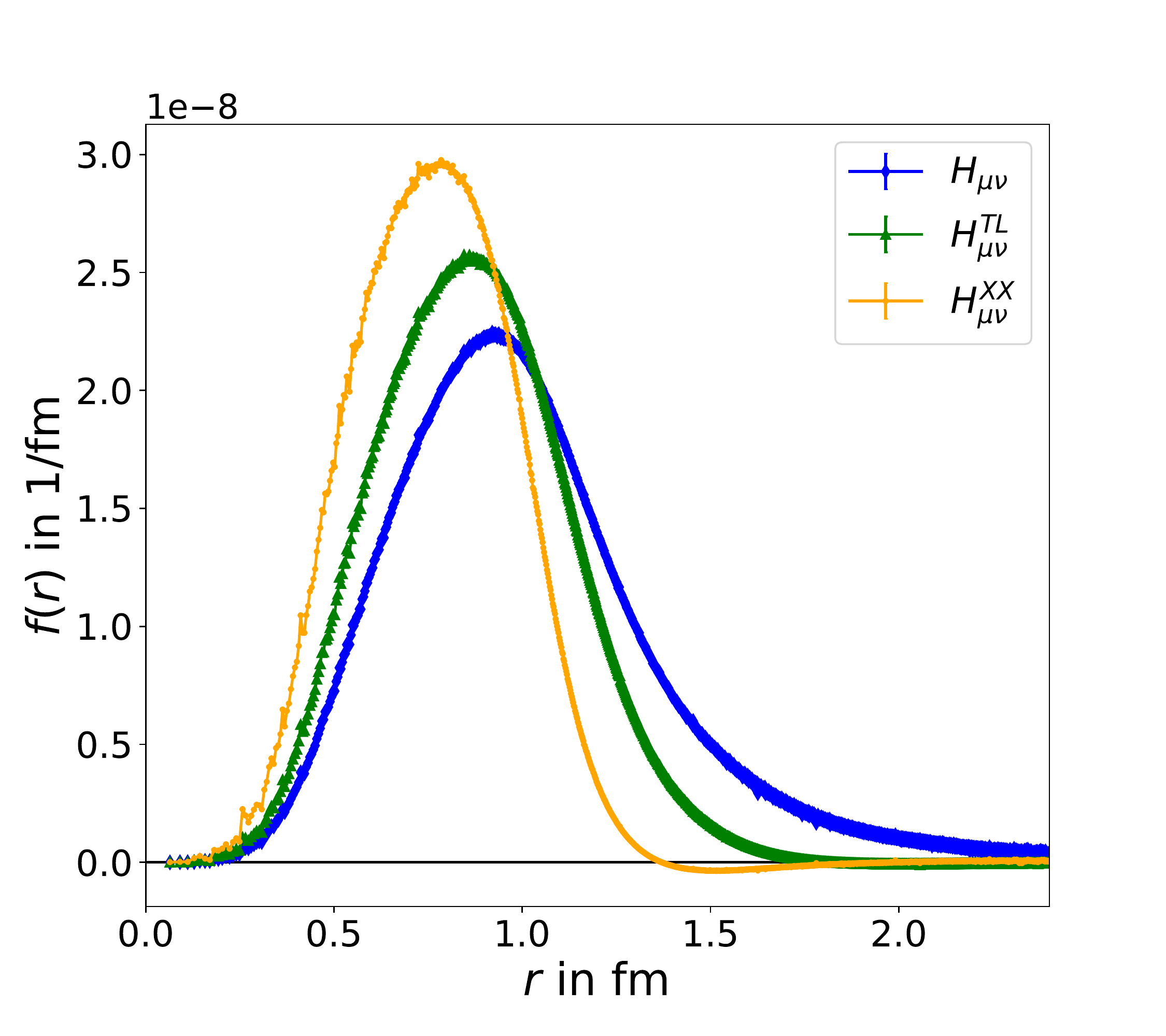}
    \caption{N203 (L=3.1fm)}
    \label{fig:N203_dif_kernels}
\end{subfigure}
\begin{subfigure}{0.49\textwidth}
     \centering
    \includegraphics[width=0.9\textwidth]{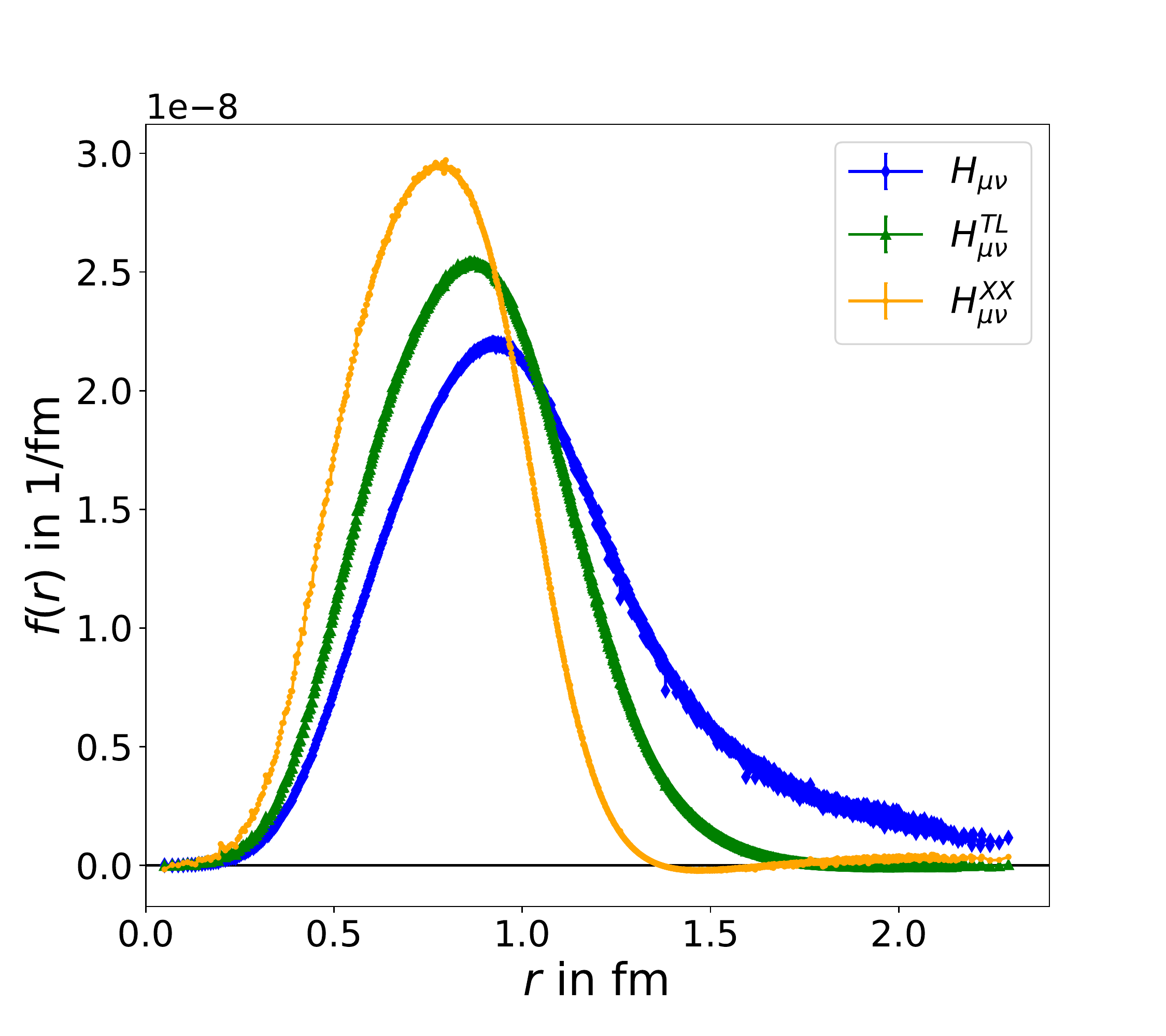}
    \caption{N302 (L=2.4fm)}
    \label{fig:N302_dif_kernels}
\end{subfigure}
	\caption{Comparison of the integrand of Eq.~\eqref{eq:awlcc} for different kernels [Eqs.~(\ref{eq:ccs-kernel},\ref{TracelessKernel},\ref{XXKernel})], for two different ensembles for the conserved-local discretization.}
\label{fig:ccs_integrand}
\end{figure}

We neglect the effects of having a finite temporal extent, as $m_\pi T$ is large for the ensembles included in this calculation.
In the CCS method, one has to correct for the FSEs coming from two sources.
The first one is the truncation of the integrand of Eq.~\eqref{eq:awlcc} at $r_{\textrm{max}}=L/2$ because of the finite lattice size. 
The resulting missing contribution could be large if the integrand is long-ranged.
Selected raw lattice results obtained with different kernels are displayed in Fig.~\ref{fig:ccs_integrand}. 
We see that the widths of the integrand are very different according to the kernel used. 
For the kernels $H_{\mu\nu}^{\textrm{TL}}$ and $H_{\mu\nu}^{\textrm{XX}}$, the integrals to get $\ahvp$ saturate more rapidly than in the case of the original, un-subtracted kernel $H_{\mu\nu}$; the FSE corrections due to the truncation are thus much smaller for the first two.

The second source of FSEs is the wrap-around-the-world effect related to the discretized momenta in a finite, periodic box.
We estimate this effect by directly comparing the correlators computed in finite- and infinite-volume Sakurai QFT [Eq.~\eqref{fse_sakurai_winding_formula}].
The finite-volume part of the latter is to be done following the same summation schemes described in Sect.~\ref{sect:summation} for different spacetime regions to match the lattice QCD calculation.
As ultimately, the relevant quantities for the calculation of $\aw$ are the contracted correlators [Eqs.~(\ref{eq:awlccg1},\ref{eq:awlccg2})], we compute the contracted finite-volume correlators at a distance $|x|=r$ by sampling them at several points $x$ equally-distributed on the same hypersphere in order to reduce the computational cost.

The numerical error of this sampling procedure is quantified based on the variation of the correction when increasing the density of the sampled points.
With our setup, we estimate the wrap-around-the-world effect to be controlled at the 10\%-level.
An additional uncertainty comes from the fact that the winding expansion Eq.~\eqref{fse_sakurai_winding_formula} is truncated at a given order.
Our choice is to truncate at $\left\Vert n \right\Vert_2^2 = 4$ and $\left\Vert n \right\Vert_2^2 + \left\Vert \nu \right\Vert_2^2 = 3$ in the first and the second sum in Eq.~\eqref{fse_sakurai_winding_formula} respectively.
An estimate of the upper bound for the truncation error is given by the highest-order kept term.
This error is added in quadrature to the uncertainty of the sampling procedure, which gives the total numerical error of the calculation. The FSE corrections computed according to the procedure described above are summarized in Tab.~\ref{table:fs_correction}.

To get an idea of the size of the systematic error associated with the use of the Sakurai QFT, we also compute the same quantity in leading-order ChPT, where the photon-two-pions coupling is described by scalar QED.
There are significant relative differences between the estimates, though the order of magnitude remains the same.
Thus we decide to quote 25\% of the total FSE correction as a modelling error, which we add in quadrature 
to the numerical error discussed in the previous paragraph.

\subsection{Comparison of the prediction for the finite-size error between the Sakurai QFT and lattice data}
\begin{figure}
\begin{subfigure}{0.49\textwidth}
     \centering
    \includegraphics[width=0.99\textwidth]{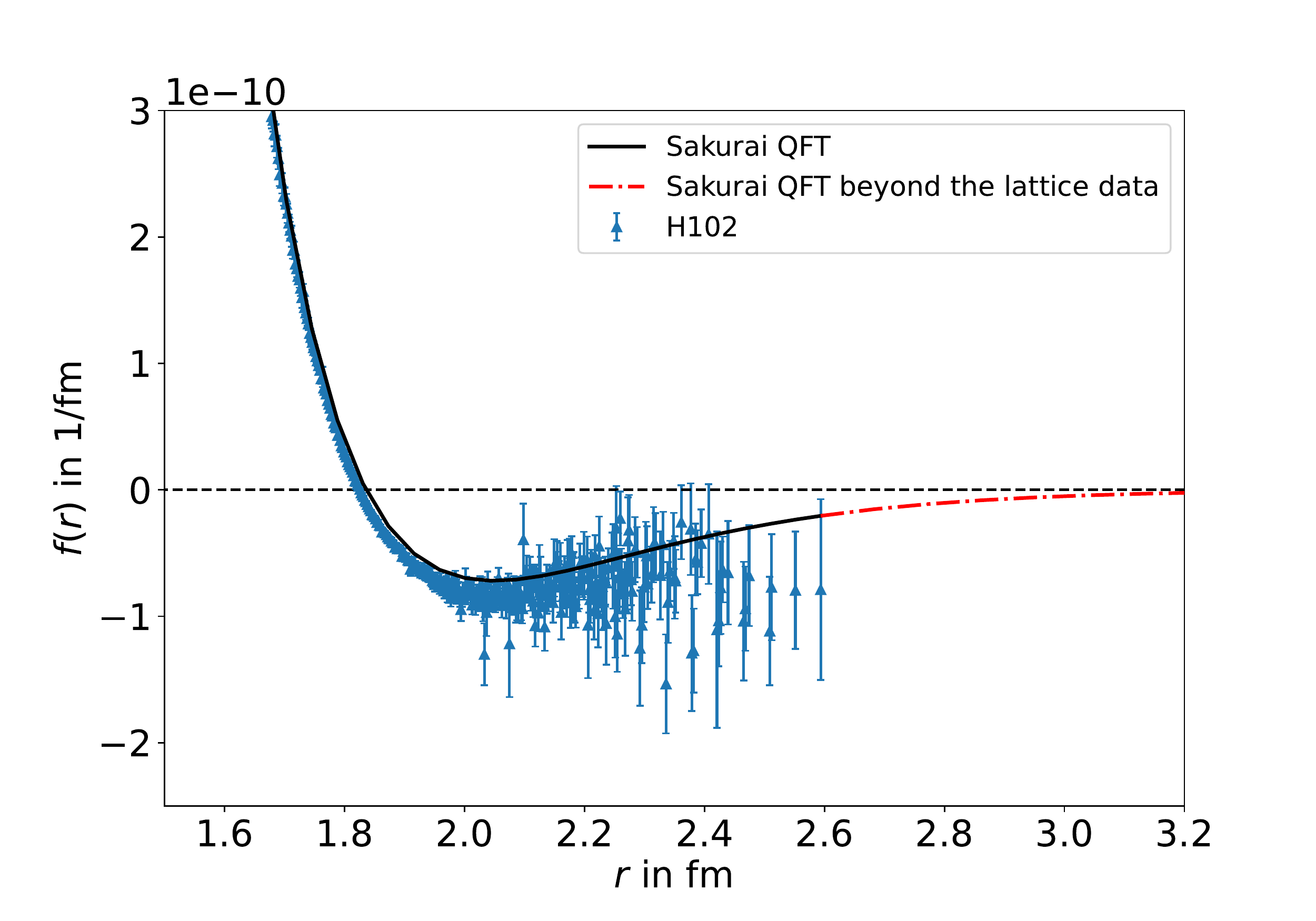}
\end{subfigure}
\begin{subfigure}{0.49\textwidth}
     \centering
    \includegraphics[width=0.99\textwidth]{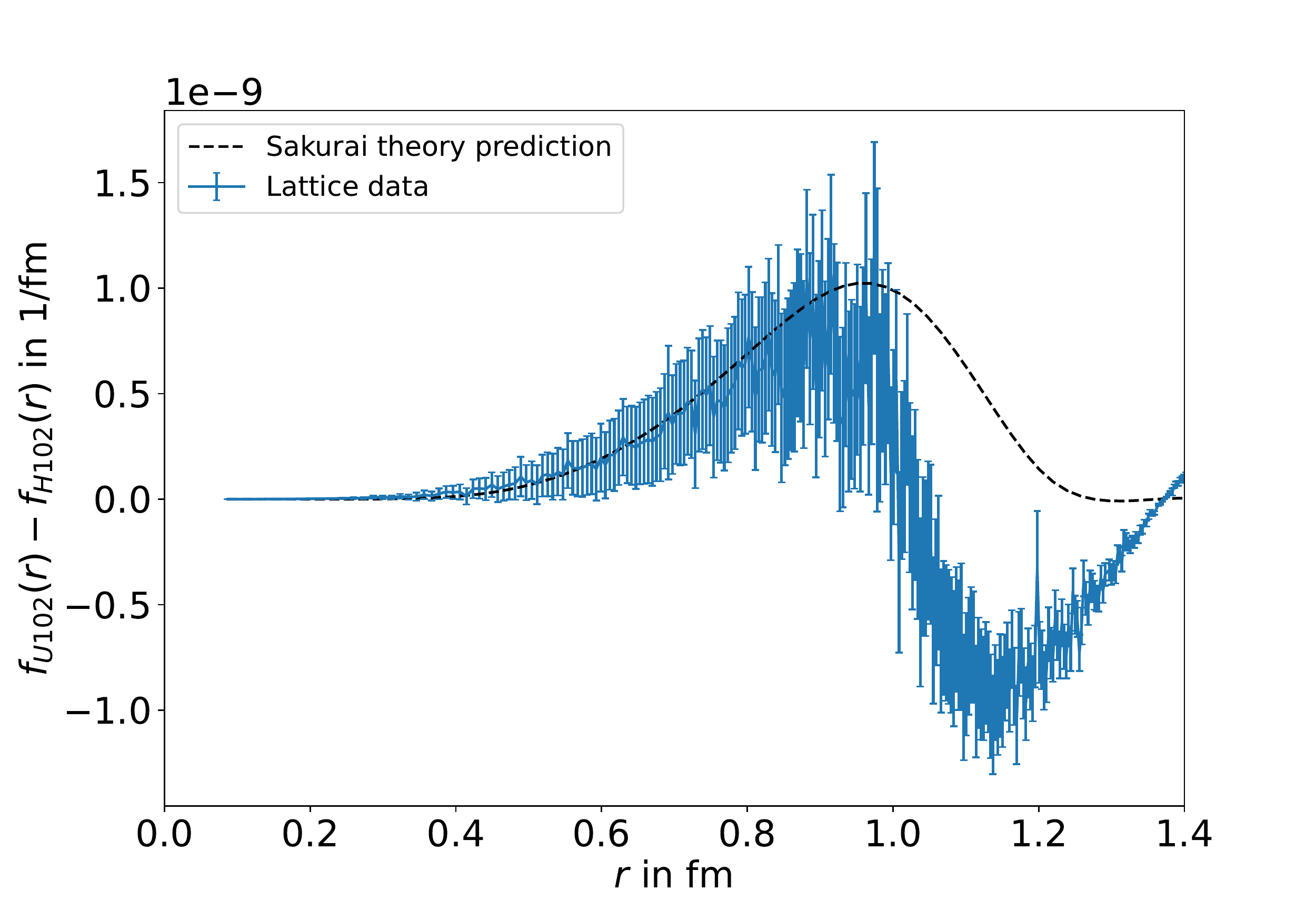}
\end{subfigure}
	\caption{Left: Plot of the lattice data from H102 and the prediction of the Sakurai QFT for the tail of the integrand. The red curve is used to calculate the correction for truncating the integrand. Right: Comparison of the difference between the ensembles U102 and H102 and the prediction from the Sakurai QFT.}
\label{fig:fs_trunc_UH}
\end{figure}

Although in Fig.~\ref{fig:ccs_integrand}, the shorter-range $H_{\mu\nu}^{\textrm{XX}}$ might appear to be beneficial in terms of its noise-to-signal ratio, we still prefer the $H_{\mu\nu}^{\textrm{TL}}$ kernel in this study for two reasons.
First, on coarser ensembles, the integrand exhibits noticeable oscillations at short distances, which indicates that the discretization effect due to the breaking of the O$(4)$-symmetry might be less well handled by performing the angular average over the available lattice points.
This effect can be observed in the comparison between the data from a coarser (N203) and finer (N302) ensemble plotted in Fig.~\ref{fig:ccs_integrand}.
The second reason for preferring the TL-kernel is that, even though the tail is strongly suppressed, the Sakurai theory still predicts non-negligible contributions in this region, if the box size is not big enough.
On the left panel of Fig.~\ref{fig:fs_trunc_UH}, we show a zoomed-in version of the tail of the integrand of H102 with the TL-kernel.
With this choice of kernel, the integrand is very well described by the Sakurai QFT.
On the other hand, with the quality of our data, using the XX-kernel in this region gives a noisy result consistent with zero, making it hard to really conclude if the model describes the long distance behavior of the integrand correctly.
Therefore, we deem it most appropriate to opt for the traceless kernel $H_{\mu\nu}^{\textrm{TL}}$ in our calculation for $\aw$, as the FSE due to the truncation seems to be better controlled.
However, one should not exclude the possibility that the shorter ranged XX-kernel might become a better choice, if only fine enough ensembles are included in the continuum extrapolation, with well-resolved tails of the integrand.

In order to test to what extent our FSE correction procedure works, we compare the difference between the integrand data computed with H102 and U102, differing only in their spatial length $L$, to the Sakurai QFT prediction at the corresponding volumes, as shown on the right panel of Fig.~\ref{fig:fs_trunc_UH}.
For this study, we set $m_\rho$ for U102 to be the same as that of H102, as only the latter is available from Ref.~\cite{gerardin:2019rua}.
The error on the lattice data is obtained by adding the statistical errors from each individual ensemble in quadrature.
Although the fluctuations on the lattice data are large compared to the central values, the prediction from the Sakurai QFT seems to follow the trend very nicely and gives the right order of magnitude up to about $r=0.8$\;fm, where the integrand from the Sakurai QFT peaks. 
However, beyond this region, the Sakurai QFT is no longer in good agreement with the lattice data.
Beside a possible mistuning in $m_\rho$ for U102, another reason for this discrepancy might be that, as we approach or go beyond the half of the linear box size (1.05 fm for U102), the convergence of the winding expansion Eq.~\eqref{fse_sakurai_winding_formula} is not sufficiently good  for such a small box.
As the summation scheme for the region beyond $r=1.05$\;fm requires one to sample the two boxes in different ways for geometrical reasons, a more careful discussion of the validity of the Sakurai QFT would be needed, especially on smaller boxes where the sensitivity of the model at short distances becomes critical.

The study described above suggests that the Sakurai QFT is able to effectively model the FSE due to the wrap-around-the-world effect of the pion up to medium values of $r$, but this effect might become too large to control with smaller boxes. 
Moreover, the correction needed to reconstruct the tail is sizeable for a small box like U102, leading to a less predictive result. 
For these reasons, the ensemble U102 is not included in the final analysis of this work.

%% file: result.tex
\section{Numerical results}\label{sect:result}
In this section, we discuss the numerical results for $\aw$ from our lattice simulations, which are based on the kernel $H_{\mu\nu}^{\textrm{TL}}$ and on the finite-size corrections detailed in Sect.~\ref{sect:fse_correction}.
We first compare the results from each individual ensemble to what has been obtained in the previous Mainz publication based on the TMR~\cite{Ce:2022kxy}.
Then, we correct for the mistuning of the pion mass to shift to the reference pion mass of 350 MeV and kaon mass of 450 MeV prior to extrapolating the data to the continuum limit.

\input{comp_tmr}

\input{extrap}

%% file: comp_tmr.tex
\subsection{Comparison to the time-momentum representation result}
\label{comparison_ensembles}
The ensemble-by-ensemble results for the isovector and strange contributions are displayed in table \ref{table:results_ensemble}.
Recall that two discretizations of the current-current correlator, namely the local-local (LL) and the conserved-local (CL), have been used to check for discretization effects (cf. Sect.~\ref{sect:lattice_setup}).
Due to the different discretization schemes, the results from this study based on the CCS method do not necessarily agree with those obtained with the TMR method.
For the strange-quark contribution, the results obtained from both methods agree with each other quite well.
Note that we do not apply any FSE correction to the strange data, as they receive contributions from the kaon loop at the leading order in ChPT, which is far more suppressed at large distances due to the higher mass of the kaon.
In the isovector channel, we observe a good agreement between the CCS and the TMR methods for the local-local data on the larger ensembles H102 and N203. 
For the smaller ensembles S400 and N302 the agreement for the local-local discretization is slightly worse.
When we look at the strange data, the agreement on the smaller ensembles is better. This could be a sign that the worse agreement in the isovector channel for S400 and N302 is due to finite-size effects, because these effects are much smaller for the strange channel.
On the contrary, for the conserved-local data we see a different behaviour, when we compare the individual ensembles: our results with the CCS method lie below the TMR values. 
This fact is a hint that the results for the conserved-local discretization show a much flatter gradient as the continuum limit is approached, since in both methods, the O($a$)-improvement has been implemented.
This behaviour is illustrated in Fig. \ref{fig:continuum_extrapolation}, when we later perform the continuum extrapolation at the common reference point.

\begin{table}[]

\begin{tabular}{|c|cccc|cccc|}
	\toprule
	& \multicolumn{4}{c|}{CCS method $H_{\mu \nu}^{\textrm{TL}}$ kernel}             & \multicolumn{4}{c|}{TMR method}                 \\
	\hline
     & \multicolumn{2}{c}{isovector} & \multicolumn{2}{c|}{strange} & \multicolumn{2}{c}{isovector} & \multicolumn{2}{c|}{strange} \\
Id   & (LL)           & (CL)         & (LL)          & (CL)         & (LL)          & (CL)          & (LL)         & (CL)         \\ \hline
U102 &  174.26(191)             &     164.78(190)         &  ---   & ---   & ---   & ---  & ---  &  ---   \\ \hline
H102 &  177.83(92)             &     168.66(90)         & 35.66(19)     & 33.54(19)    &  178.54(52)   &   179.75(52)  & 35.66(12)    & 35.90(11)    \\ \hline
S400 &    175.21(96)           &    167.57(94)        & 34.90(20)     & 33.15(20)    &  173.82(69)   &   174.49(68)  & 34.402(86)   & 34.548(82)   \\ \hline
N203 &     173.25(89)           &    167.60(88)         & 34.11(14)     & 32.83(13)    &   173.75(43)  &  174.11(43)   & 34.225(90)   & 34.283(89)   \\ \hline
N302 &    169.08(96)          &       165.39(95)       & 33.31(17)     & 32.46(17)    &  167.77(87)   &  167.84(87)   & 32.427(83)   & 32.444(82)   \\ \botrule
\end{tabular}
	\caption{Comparison between the results for the isovector and strange connected contribution obtained in the CCS method using spherical integration and the results of the Mainz group \cite{Ce:2022kxy} using the TMR method. Finite size corrections are applied to the isovector contribution for both methods. The results for U102 are not included in the final analysis. All values are in units of $10^{-10}$.}
\label{table:results_ensemble}
\end{table}

\subsection{Shift to a common reference point}
\label{sect:shift}
The chosen ensembles from table \ref{table:ensemble} are not exactly at the same pion and kaon mass. 
Although these masses are not very different, we want to shift the results for each ensemble to a common reference point in the $(m_\pi, m_K)$-phase-space.
We define this reference point to be at $m_\pi=350$ MeV and $m_K=450$ MeV.
For this task, we use one of the best global fits from the calculation of the Mainz group in the TMR method~\cite{Ce:2022kxy}. For the isovector contribution the fit has the following form
\be
\label{eq:tmr_fit_l1}
\begin{split}
(a_\mu^{\textrm{W}})_{I=1}(a,\phi_2,\phi_4) =&\quad p_0 + p_1 (\phi _2 - \phi _{2,{\mathrm{phys}}}) + p_2 (\log(\phi_2) - \log(\phi_{2,{\mathrm{phys}}})) \\
	&  + p_3 (\phi_4 - \phi_{4,{\mathrm{phys}}}) + p_4 a^2\,,
\end{split}
\ee
and for the strange contribution we have
\be
\label{eq:tmr_fit_s}
\begin{split}
(a_\mu^{\textrm{W}})_{\textrm{strange}}(a,\phi_2,\phi_4) =&\quad p_0 + p_1 (\phi _2 - \phi _{2,{\mathrm{phys}}}) + p_2 (\phi_2 - \phi_{2,{\mathrm{phys}}}) \\
	&  + p_3 (\phi_4 - \phi_{4,{\mathrm{phys}}}) + p_4 a^2\,.
\end{split}
\ee
The fit parameters $p_i$ and the associated covariance matrices are taken from the calculation done in Ref. \cite{Ce:2022kxy}. 
In the above, $a$ is the lattice spacing, $\phi_2\equiv 8t_0 m_\pi^2$ and $\phi_4\equiv 8t_0(m_K^2 + \frac{1}{2}m_\pi^2)$ are the dimensionless parameters defined with the gradient flow time $t_0$~\cite{Luscher:2010iy}.
With this fit form we calculate the differences between the result at the reference point and the result at the pion and kaon mass of the specific ensemble. 
This difference is independent of the  lattice spacing of the given ensemble. 
The errors are calculated from the covariance matrices of the fits and the results of this calculation are given in tab.~\ref{table:corrections_to_ref}. 
We then apply these differences as a correction to the results on each ensemble in the CCS method.
We used the TMR fit for the same current discretization (LL, CL) to correct the corresponding CCS data. 
However, we see that there is only a very small difference between the shifts for the LL and CL discretization calculated in the TMR method. 

Again, $25\%$ of the correction is assigned for the systematic uncertainty for this procedure. 
Since the chosen ensembles are very close to the chosen reference point, the systematic errors from shifting to that reference point are very small.

%% file: extrap.tex
\subsection{Continuum extrapolation}

\begin{figure}
\begin{subfigure}{0.49\textwidth}
    \centering
    \includegraphics[width=0.9\textwidth]{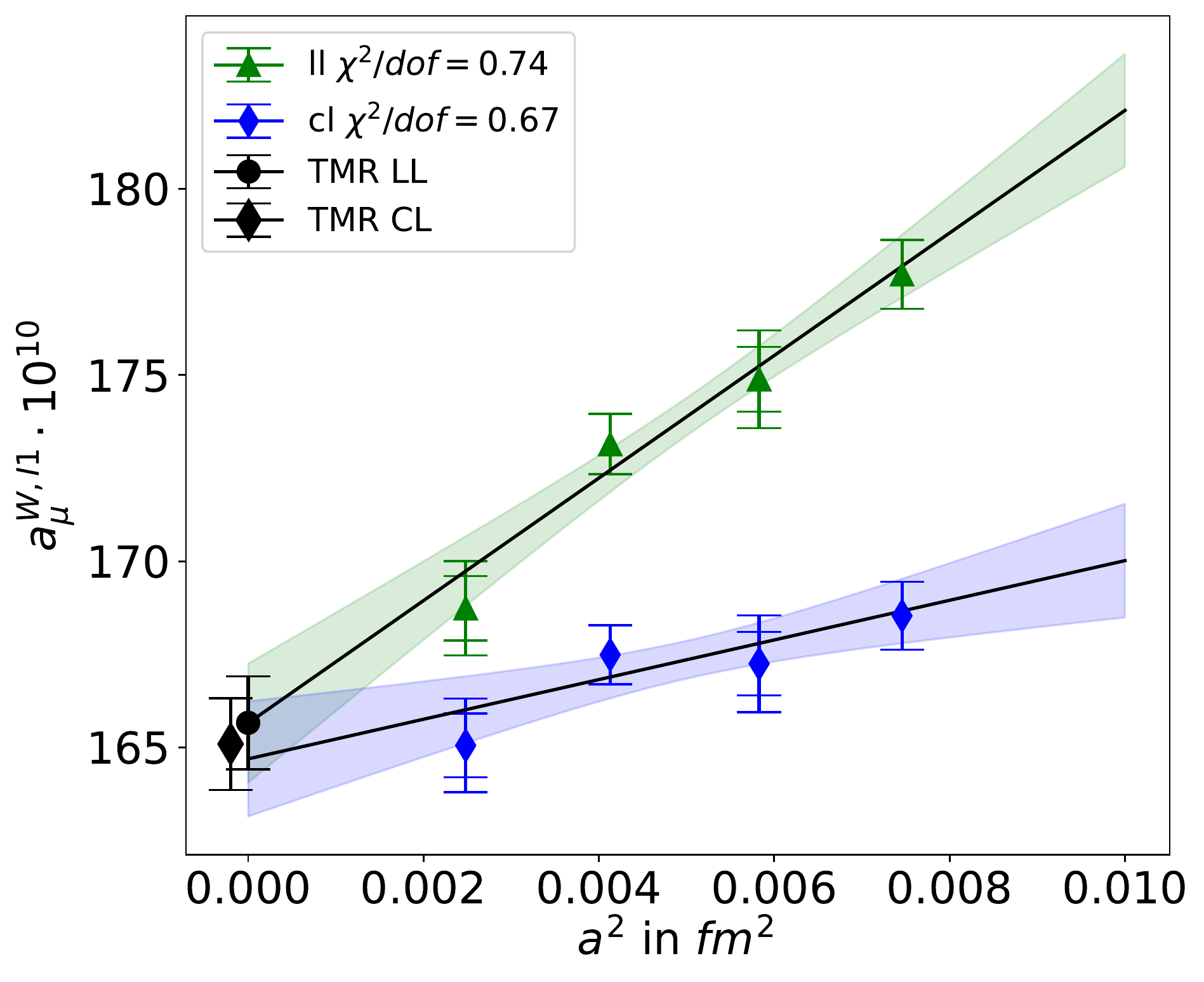}
    \caption{Isovector contribution}
\end{subfigure}
\begin{subfigure}{0.49\textwidth}
    \centering
    \includegraphics[width=0.9\textwidth]{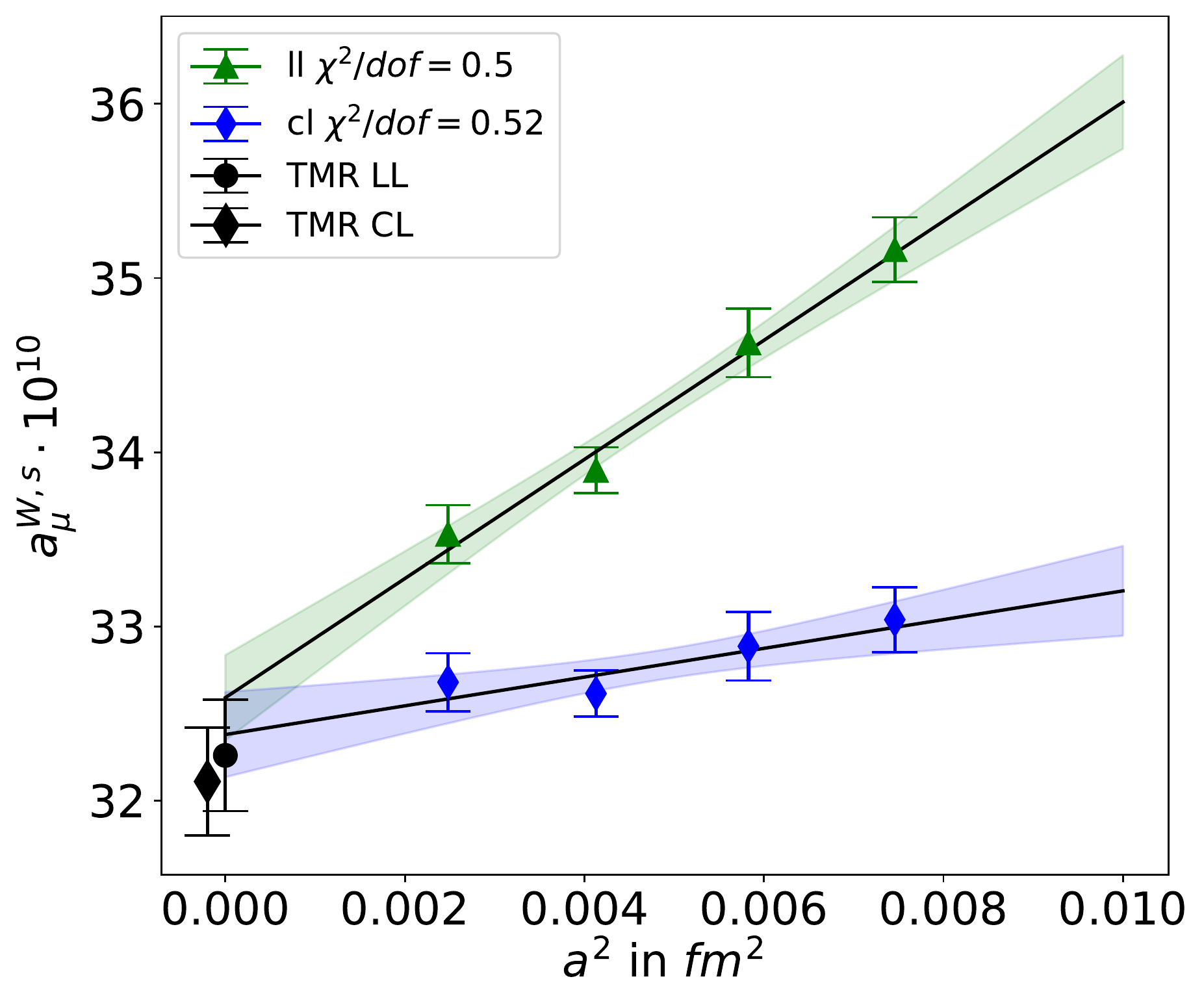}
    \caption{Strange contribution }
\end{subfigure}
	\caption{Continuum extrapolation at the reference point $m_\pi=350$ MeV and $m_K = 450$ MeV using the TL-kernel. The results from the TMR method are both at $a=0$. They are separated slightly for a better visibility. The isovector contribution is corrected for finite-size effects. For the strange contribution no finite-size correction is applied. The smaller error bar is only the statistical error, the larger is the total error. The systematic error on N203 and H102 in the isovector contribution is almost not visible. For the strange contribution the uncertainty on each ensemble is highly dominated by the statistical error, as the systematic error from the shift to the reference point is not visible.}
\label{fig:continuum_extrapolation}
\end{figure}

After we applied the corrections to account for the mistuning of the pion masses to the reference point, we perform an extrapolation to the continuum with a linear fit in $a^2$
\begin{equation}\label{eq:fit_f1}
f_1(a,\alpha_1,\beta_1) = \alpha_1 + \beta_1 a^2\,.
\end{equation}
This is depicted in Fig.~\ref{fig:continuum_extrapolation} and the results of the continuum extrapolation are displayed in Tab.~\ref{table:continuum_result}. 

\begin{table}[]
	\begin{tabular}{|c|cccc|}
	\toprule
& \multicolumn{2}{c}{isovector} & \multicolumn{2}{c|}{strange} \\
\hline
 Id                        & (LL)          & (CL)          & (LL)         & (CL)         \\ \hline
$H_{\mu\nu}^{\textrm{TL}}$ & 165.75(158) & 164.69(156)  & 32.61(24)  & 32.38(23)  \\ \hline
TMR \cite{Ce:2022kxy}                & 165.66(125)   & 165.09(123)   & 32.26(32)    & 32.11(31)  \\ 
\botrule
\end{tabular}
	\caption{Results of the continuum extrapolation from the CCS method and the TMR method with statistical uncertainties. The results of the TMR method are obtained from the fits in Eqs.~\eqref{eq:tmr_fit_l1} and \eqref{eq:tmr_fit_s}.
          All values are in units of $10^{-10}$.
 }
\label{table:continuum_result}
\end{table}

\begin{figure}
\begin{subfigure}{0.576\textwidth}
    \centering
    \includegraphics[width=0.99\textwidth]{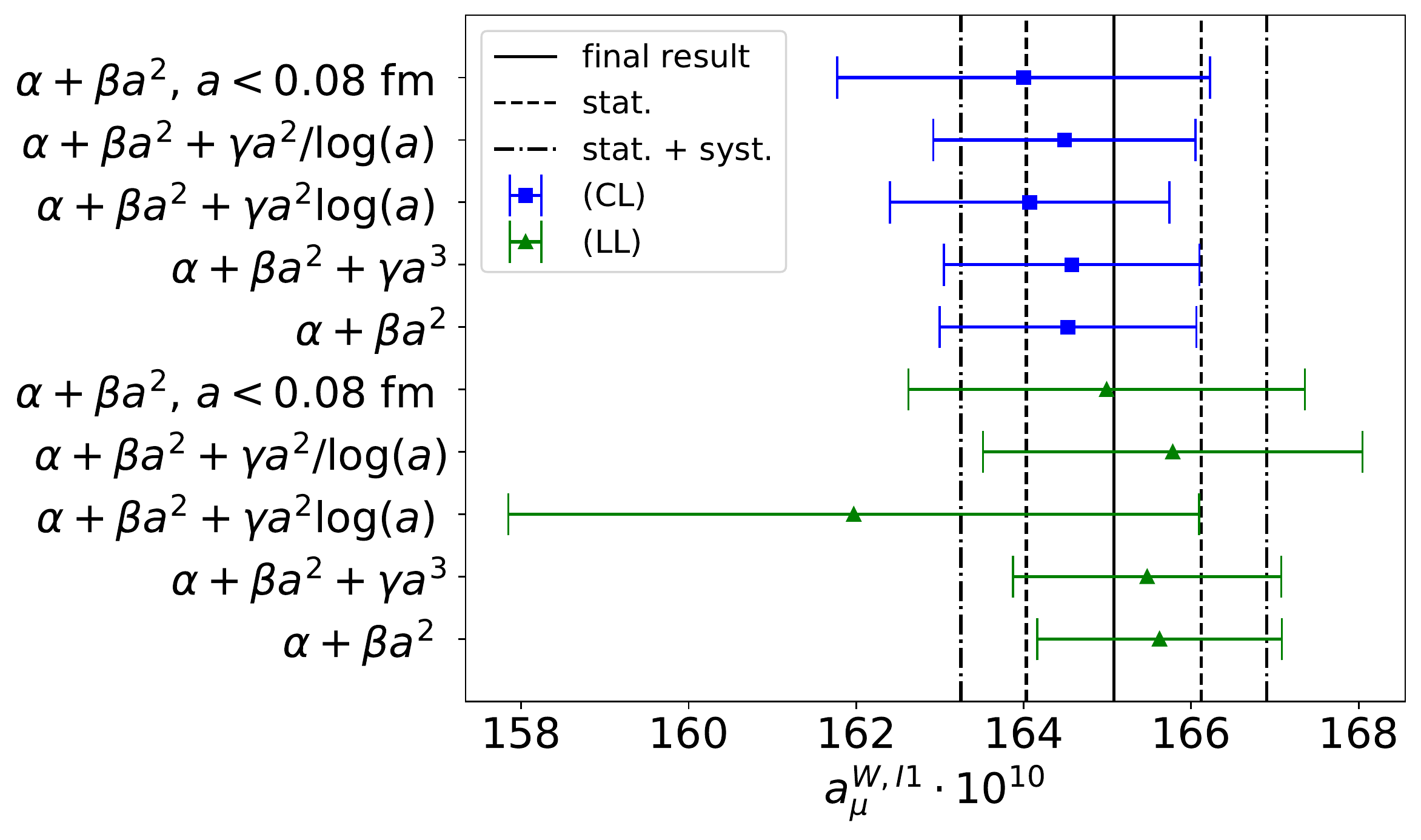}
    \caption{Isovector contribution}
\end{subfigure}
\begin{subfigure}{0.411\textwidth}
    \centering
    \includegraphics[width=0.99\textwidth]{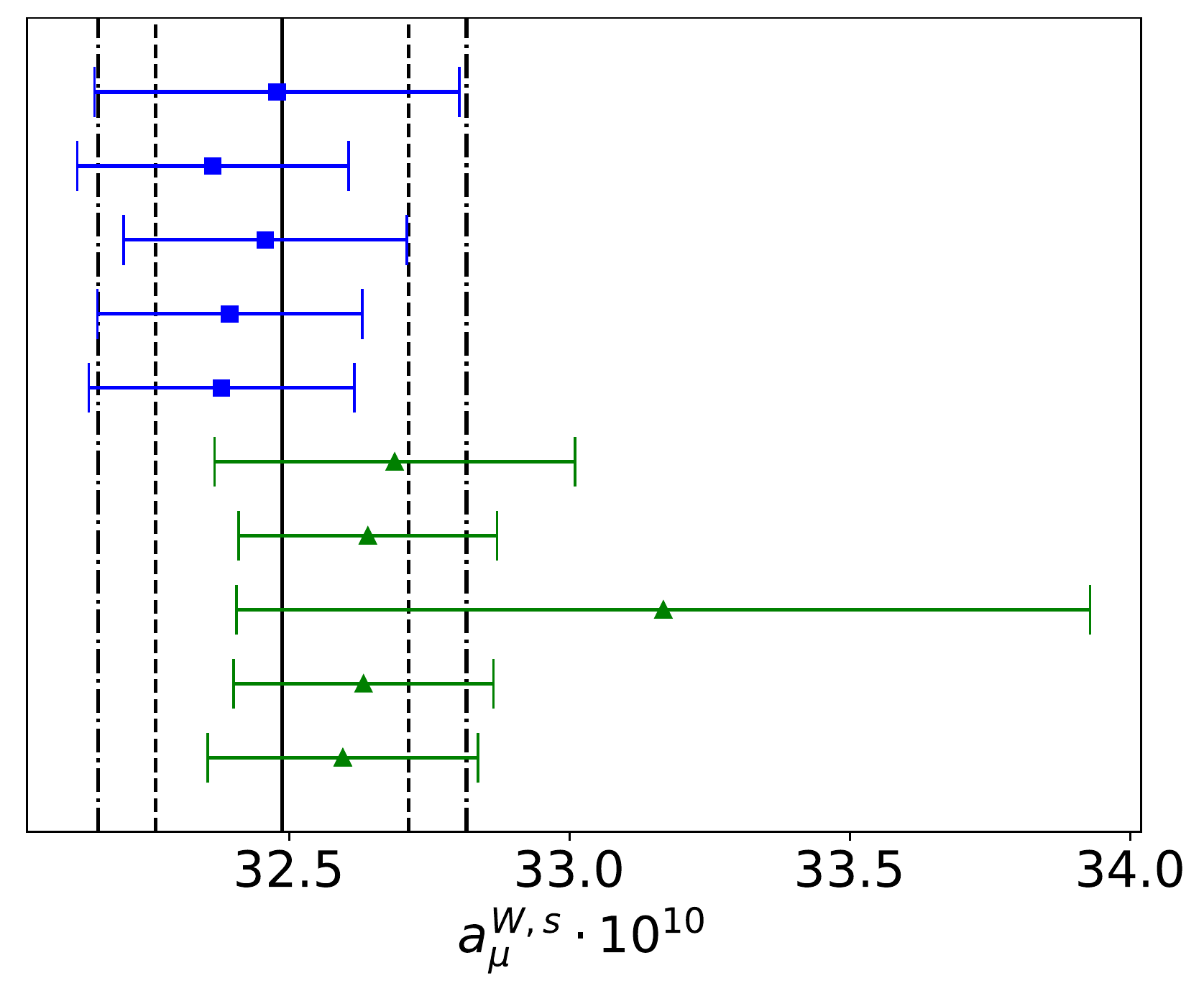}
    \caption{Strange contribution }
\end{subfigure}
	\caption{Comparison of the different fit-ansätze for the continuum extrapolation at the reference point $m_\pi=350$ MeV and $m_K = 450$ MeV using the TL-kernel. The root-mean-square deviation of all the different fits is calculated and gives the systematic uncertainty of the continuum extrapolation.}
\label{fig:cont_syst}
\end{figure}

Since the O(a)-improvement procedure is fully implemented, O(a) artifacts are expected to be absent in the continuum extrapolation. 
However, higher order terms, such as $a^3$, $a^2\log(a)$ and $a^2/\log(a)$ could also be non-negligible. 
This leads to a systematic error of the extrapolation. 
In order to obtain an estimate of this uncertainty, we perform several additional fits.
For each of the fits, we allow one of these terms to be non zero. 
This makes us consider the following additional three-parameter fit-ans\"atze
\ba 
\label{eq:fit_f2}
f_2(a,\alpha_2,\beta_2,\gamma_2) &=& \alpha_2 + \beta_2 a^2 +\gamma_2 a^3\,,\\
\label{eq:fit_f3}
f_3(a,\alpha_3,\beta_3,\gamma_3) &=& \alpha_3 + \beta_3 a^2 + \gamma_3 a^2 \log(a)\,,\\
\label{eq:fit_f4}
f_4(a,\alpha_4,\beta_4,\gamma_4) &=& \alpha_4 + \beta_4 a^2+ \gamma_4 \frac{a^2}{\log(a)}.
\ea 
These fit ans\"atze leave only one degree of freedom with our available data. Hence, over-fitting could potentially be an issue.
We observe a large cancellation between the term multiplying $\beta_i$ and the one multiplying $\gamma_i$. 
Lacking guidance from additional data points, we introduce Gaussian priors to constrain the highest order terms in $a$, $\gamma_i$, in the ans\"atze Eqs.~(\ref{eq:fit_f2}-\ref{eq:fit_f4}) to be in similar size as the best-fit coefficient $\beta_1$ from Eq.~\eqref{eq:fit_f1}.
Additionally, to probe the sensitivity of the linear fit $f_1$ to the range in lattice spacing of the data, we also perform the fit with the coarsest lattice spacing left out.
We apply this procedure to the LL and CL data independently, resulting in 10 different fits. 
To get an estimate of the systematic error of the fitting procedure, we calculate the root-mean-squared deviation of the individual fit results in the continuum limit $y_i$ from their average $\bar{y}$, $\Delta y_{\textrm{RMS}} \equiv \Big(\sum_{i=1}^N(y_i-\bar{y})^2/N \Big)^{1/2}$.
The results for $\aw$ from the different fits are shown in Fig.~\ref{fig:cont_syst}.
We see that the extrapolations for the conserved and the local current are in good agreement. 
Furthermore, the continuum values at the reference point are consistent with the calculation with the TMR method.
\\

For our final estimate for the isovector and the strange-quark contribution to $\aw$ with the CCS method at the reference point of $m_\pi=350$ MeV and $m_K=450$ MeV, we quote the result from a constant fit to the LL and CL outcomes under the fit-ansatz $f_1$:
\ba
a_\mu^{\textrm{W,I1}} &=&  165.17 (157)_{\textrm{stat}}(99)_{\textrm{syst}}\times 10^{-10}\,,\\
a_\mu^{\textrm{W,s}} &=&  32.49(22)_{\textrm{stat}}(23)_{\textrm{syst}}\times 10^{-10}\,.
\ea

%% file: conclu.tex
\section{Conclusion}\label{sect:conclu}
In this work, we have extended the Covariant Coordinate Space method first proposed in Ref.~\cite{Meyer:2017hjv} to the window quantity for the anomalous magnetic moment of the muon.
Due to the stark geometric difference to the Time-Momentum Representation, this alternative approach provides a valuable cross-check for the existing window quantity results from Lattice QCD.
We provide values for the intermediate window quantity in the isovector channel and for the strange quark-connected contribution at $m_\pi=350$ MeV and $m_K=450$ MeV.
With an appropriate finite-size effect correction scheme and a careful scrutiny of the discretization effects, we obtain $\aw=165.17(186)\times 10^{-10}$ for the isovector contribution and $\aw=32.49(32)\times 10^{-10}$ for the strange quark-connected contribution, where the statistical and systematic errors have been added in quadrature, confirming the results of the calculation of the Mainz group using the TMR method~\cite{Ce:2022kxy}.
This study strengthens the tension between the lattice calculations and the dispersive approach on the window quantity.

One advantage of the CCS method is the freedom to modify the weight of the correlator computed at different regions without changing the final summed answer. This might turn out useful especially if one wants to adjust the shape of the lattice integrand to mitigate statistically noisy contributions. Future applications might involve different weight functions for different Euclidean time windows to optimize the integrand for minimal lattice artifacts and statistical noise.
Furthermore, we have demonstrated how to correct for the finite-size effects in the CCS method based on an effective field theory approach.
A strong motivation for this strategy is the non-trivial symmetric rank-two tensor structure of the coordinate-space correlator required by the formalism.
A simple $\rho$-$\gamma$ mixing model advocated by Jegerlehner and Szafron~\cite{Jegerlehner:2011ti} successfully captures the long-distance contribution to $\aw$ in the CCS representation. 
The expected $a^2$-scaling that our data shows after the finite-size correction based on this model is encouraging and suggests that the same model might also be utilized as a guideline for further optimizations with the CCS method.
This is of special interest for the calculation of the full Hadronic Vacuum Polarization to $a_\mu$, whose integrand is much longer-ranged than $\aw$.
The technical details appended to this paper might be useful while computing other coordinate-space observables with similar integrable divergences in momentum-space.

The present calculation can be easily carried over to calculations of other lattice observables such as the full Hadronic Vacuum Polarization contribution to $a_\mu$ or the running of the QED coupling.
For these observables, it might be of interest to combine the CCS method with master field simulations~\cite{Fritzsch:2021klm}.
These simulations are performed over very large lattices, thus finite-size effects are expected to be highly suppressed.
In particular, we expect that this framework is the best suited for studying the quark-disconnected contribution, which has been omitted in this work.
It might be possible to get a more precise determination of this contribution with a short-ranged CCS kernel to filter out the noisy region for lattice calculations.

%% file: window_kernel.tex
\section{Derivation of the kernel for the window quantity in the CCS representation}
\label{appendix:window_kernel}

Let $G(t)$ as defined in Eq.~\eqref{TMRcorrelator} be the (positive-definite) TMR correlator.
The relation to the vacuum polarization function\footnote{The HVP function is defined as
in Ref.\ \cite{Bernecker:2011gh}.} is 
\be\la{eq:Gt}
G(t) \stackrel{t\neq 0}{=} \int_{-\infty}^\infty \frac{d\omega}{2\pi} \, \omega^2 \, [\Pi(\omega^2)-\Pi(0)] \, e^{i\omega t}.
\ee
Introducing the Adler function
\be
{\cal A}(\omega^2 ) = \omega^2 \, \frac{d}{d\omega^2}\Pi(\omega^2),
\ee
one obtains after writing
\be
\Pi(\omega^2 ) -\Pi(0) = \int_0^{\omega^2} \frac{ds}{s}\,{\cal A}(s) 
\ee
and integrating by parts over $\omega$ (i.e. $e^{i\omega t} = \frac{d}{d\omega}\frac{e^{i\omega t}}{it}$)
in Eq.~(\ref{eq:Gt}),
\be\la{eq:Gt2}
G(t) = \frac{1}{\pi} \int_0^\infty \frac{d\omega^2}{\omega^2}\,{\cal A}(\omega^2)\; \frac{d^2}{dt^2}\left(\frac{\sin (\omega t)}{t}\right).
\ee

Let now an observable in the TMR be given by
\be
a_\mu^W = \int_0^\infty dt \;f_W(t)\,G(t).
\ee
Inserting expression (\ref{eq:Gt2}) for the correlator $G(t)$,
one finds
\be\la{eq:amuW}
a_\mu^W = \int_0^\infty dQ^2 \,{\cal A}(Q^2) \, g_W(Q^2),
\ee
with
\be
g_W(Q^2) = \frac{1}{\pi Q^2} \int_0^\infty dt\,f_W(t) \,\frac{d^2}{dt^2}\left(\frac{\sin (|Q| t)}{t}\right).
\ee
For an expression of the type (\ref{eq:amuW}), Ref.\ \cite{Meyer:2017hjv}  (Eq.~(33) therein)
gives an expression for the weight functions ${\cal H}_1$ and ${\cal H}_2$ to be used in the CCS method.
Explicitly,
\ba
 a_\mu^W &=& \int d^4x\;G_{\mu\nu}(x)\; H_{\mu\nu}(x),
\\
H_{\mu\nu}(x) &=& -\delta_{\mu\nu} {\cal H}_1(|x|) + \frac{x_\mu x_\nu}{x^2} \,{\cal H}_2(|x|),
\ea
with 
\ba
{\cal H}_i(|x|) &=& \frac{2}{3}\int_0^\infty \frac{dQ^2}{Q^2}\; h_i(|Q||x|)\, g_W(Q^2)
\\ &=& \frac{2}{3\pi}\int_0^\infty dt\,f_W(t) \,\frac{d^2}{dt^2} \left[
  \frac{1}{t} \int_0^\infty \frac{dQ^2}{Q^4}\; h_i(|Q||x|)\,  \sin (|Q| t) \right]\,.
\label{eq:Hi}
\ea
One finds, with $r=|x|$, 
\ba
\frac{1}{t}\int_0^\infty \frac{dQ^2}{Q^4}\; h_1(|Q||x|)\,  \sin (|Q| t)
&=& \frac{\theta(r-t)}{120} \, \bigg(\frac{\sqrt{r^2-t^2}
   \left(32 r^4+11 r^2 t^2+2   t^4\right)}{r^4}
\nonumber   \\ && \qquad \quad \qquad -45 t \arccos\left({t}/{r}\right)\bigg),
   \\
\frac{1}{t}   \int_0^\infty \frac{dQ^2}{Q^4}\; h_2(|Q||x|)\,  \sin (|Q| t) &=&
   \theta(r-t) \frac{\left(r^2-t^2\right)^{5/2}}{15 r^4}.
\ea
In the second derivatives, needed in Eq.~\eqref{eq:Hi}, the terms proportional to $\delta(t-r)$
or its derivative do not contribute to the ${\cal H}_i$,
as long as $f_W(t)$ is smooth. One then finds
\ba\la{eq:H1final}
{\cal H}_1(|x|) &=& \frac{2}{9\pi r^4}\int_0^r dt\, \sqrt{r^2-t^2} \left(2 r^2+t^2\right) \,f_W(t),
\\
{\cal H}_2(|x|) &=& \frac{2}{9\pi r^4} \int_0^r dt\, \sqrt{r^2-t^2}\left(4 t^2 - r^2\right) \,f_W(t)   .
\la{eq:H2final}
\ea

It is worth noting that if $f_W(t)$ practically vanishes beyond a distances $t_1$, then
for $|x|\gg t_1$,
\ba
{\cal H}_1(|x|) &\simeq & \frac{4}{9\pi |x|}\int_0^\infty dt \,f_W(t),
\\
{\cal H}_2(|x|) &\simeq & \frac{-2}{9\pi |x|} \int_0^\infty dt\, \,f_W(t)   .
\ea
Therefore, these weight functions have a long tail, unlike $f_W(t)$.
Still, the $1/|x|$ behaviour amounts to a suppression compared to the weight functions
for $a_\mu^{\rm hvp}$, which grow like $x^2$  at large $|x|$.
In the specific case of the `window quantity',
numerical integration of Eqs. (\ref{eq:H1final}--\ref{eq:H2final}) yields the weight functions displayed in Fig.\ \ref{fig:wf}.

\begin{figure}
\centerline{\includegraphics[width=0.7\textwidth]{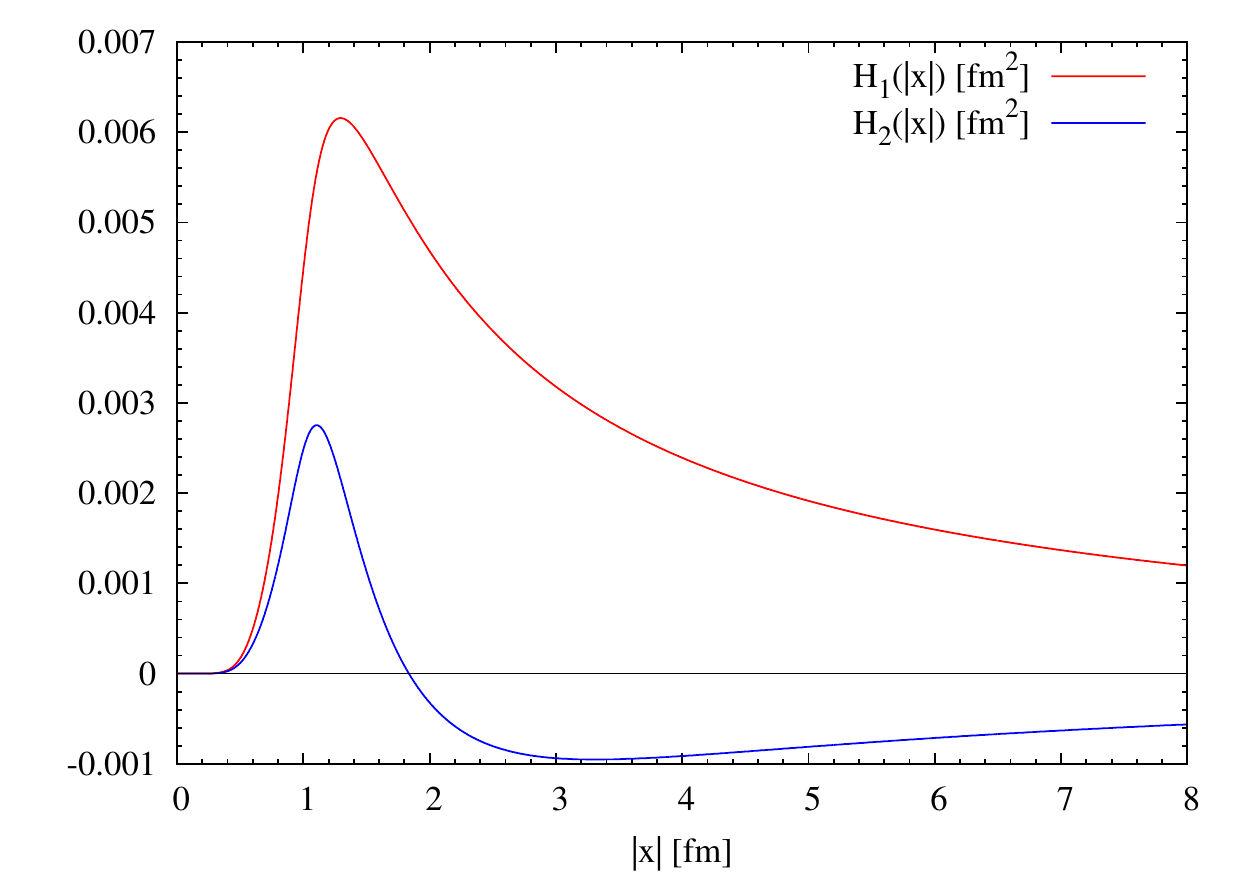}}
\caption{\la{fig:wf}
The weight functions for obtaining the intermediate window $a_\mu^W$ (defined by $t_0=0.4$\,fm, $t_1=1.0$\,fm, $\Delta=0.15$\,fm)
in the CCS method.}
\end{figure}

%% file: fse_sakurai.tex
\section{Determining the finite-size correction using Sakurai's  field theory}
\label{appendix:fse_sakurai}


In this section we discuss some features of the Sakurai QFT in details, with special focus on its renormalization to one-loop and numerical applications to the finite-size correction.
Recall that in the original basis of fields, the Euclidean spacetime Lagrangian of the theory is given in Eq.~\eqref{eq:LEinit}.
We use dimensional regularisation in the following.
Thus we are in
\be
d=2\lambda+2= 4-\varepsilon
\ee
dimensions. The massive scalar propagator reads
\be
G_m(x) = \frac{m^\lambda}{(2\pi)^{\lambda+1}}\; \frac{K_\lambda(m|x|)}{|x|^\lambda}\,
\stackrel{d=4}{=} \frac{m}{4\pi^2 |x|}K_1(m|x|),
\ee
and the massive vector propagator 
\be\label{eq:Gv}
G_{\mu\nu}(x)\equiv \<\rho_\mu(x) \rho_\nu(0)\> = \int\frac{d^dk}{(2\pi)^d} \,e^{ikx} \,\frac{\delta_{\mu\nu}+k_\mu k_\nu / m_\rho^2}{k^2+m_\rho^2}
=  \left(\delta_{\mu\nu} - \frac{1}{m_\rho^2}\partial_\mu \partial_\nu \right) G_{m_\rho}(x).
\ee

We begin by determining the couplings $g$ and $g_\gamma$, working at tree-level.
The kinematic mixing term between rho and photon can be removed at the cost of generating a direct coupling
of the rho to electrons, and it is instructive to work in this new basis.
We set
\be
\epsilon = \frac{e}{g_\gamma}
\ee
and remove the kinetic mixing term by a field transformation,
\be
\left( \begin{array}{c} A_\mu \\ \rho_\mu  \end{array} \right)
=  \left( \begin{array}{c@{\qquad}c}  1 & -\frac{\epsilon}{\sqrt{1-\epsilon^2}} \\ 0  & \frac{1}{\sqrt{1-\epsilon^2}}  \end{array} \right)
\left( \begin{array}{c} \tilde A_\mu \\ \tilde \rho_\mu  \end{array} \right)
\ee
The square-mass for the $\tilde \rho_\mu$ field is
\be
\tilde m_\rho^2 = \frac{m_\rho^2}{1-\epsilon^2}.
\ee
The covariant derivative takes the form
\ba
D_\mu &=& \partial_\mu - ie A_\mu -ig\rho_\mu
= \partial_\mu -i e \tilde A_\mu -i  \tilde g \tilde\rho_\mu,
\\
\tilde g &=& \frac{g - e\epsilon }{\sqrt{1-\epsilon^2}}.
\ea
Thus
\be
   {\cal L}_E = \frac{1}{4}F_{\mu\nu}(\tilde A)^2 + \frac{1}{4} F_{\mu\nu}(\tilde\rho)^2
   + \frac{1}{2} \tilde m_\rho^2 \tilde\rho_\mu^2 + (D_\mu\pi)^\dagger (D_\mu\pi) + m_\pi^2 \pi^\dagger \pi.
\ee
While the form of the Lagrangian is now simpler, in the new basis of
fields the electromagnetic current of the electron couples not only to
$\tilde A_\mu$, but also directly to $\tilde\rho_\mu$.
Explicitly, the coupling to the electron reads
\be
{\cal L}_E =  \bar e ((\partial_\mu -ieA_\mu) \gamma_\mu + m_e) e
= \bar e ((\partial_\mu -ie\tilde A_\mu + ie \frac{  \epsilon }{\sqrt{1-\epsilon^2}} \tilde\rho_\mu) \gamma_\mu + m_e) e
\ee
From here, one calculates the treelevel decay width for $\tilde \rho \to e^+ e^-$ by standard QFT methods (similar to
the calculation $Z^0\to \bar \ell \ell$) and finds, neglecting the electron mass,
\be
\Gamma_{e^+e^-} = \frac{1}{3} \alpha \frac{\epsilon^2}{1-\epsilon^2} \,\tilde m_\rho .
\ee
From the PDG, we set 
\be
\tilde m_\rho = 775.26(23)\,{\rm MeV}, \qquad \Gamma_{e^+e^-} = (4.72\times 10^{-5})\times 149.1\,{\rm MeV} = 7.04 \,{\rm keV},
\ee
and from here find
\be\label{eq:sakuriggamma}
\epsilon = 0.0610, \qquad g_\gamma = \frac{\sqrt{4\pi\alpha}}{\epsilon} = 4.97.
\ee
From the value of $\epsilon$, one sees that one could have dropped the factor $1/\sqrt{1-\epsilon^2}$.

Similarly, the $\tilde\rho\pi\pi$ decay is driven at leading order by the interaction $\Delta{\cal L}_E = \tilde g \tilde \rho_\mu j_\mu$.
Here, if $p$ and $q$ are respectively the final-state momenta of the $\pi^+$ and $\pi^-$, we have
\be
i{\cal M}^{(\sigma)} = \tilde g \,\epsilon^{(\sigma)}_\nu (p^\nu - q^\nu),
\ee
and
\be
\frac{1}{3}\sum_{\sigma=0,\pm} |{\cal M}^{(\sigma)} |^2 = \frac{2\tilde g^2}{3} (p\cdot q - m_\pi^2).
\ee
In the CM frame, where the norm of the pion spatial momentum is $p_\pi$, one obtains
\be
\Gamma_{\pi^+\pi^-}= \frac{p_\pi}{8\pi m_\rho^2}\frac{1}{3}\sum_{\sigma=0,\pm} |{\cal M}^{(\sigma)} |^2  = \frac{\tilde g^2 p_\pi^3}{6\pi m_\rho^2}.
\ee
Setting this equal to the experimental value of $149.1\,$MeV, one extracts
\be\label{eq:sakurig}
\tilde g = 5.976, \qquad g = 5.984.
\ee

For orientation, we note that the contribution of a narrow resonance to the $R$-ratio reads
\be
R(s) = \frac{9\pi}{\alpha^2} \, \frac{\Gamma(m_V\to e^+e^-)}{m_V} \, m_V^2 \delta(s-m_V^2).
\ee
With the expression of the $\rho$ electronic width above, ignoring the fact that it is rather broad,
this becomes
\be
R(s) = \frac{12\pi^2}{g_\gamma^2} \,  m_V^2 \delta(s-m_V^2).
\ee
Given that $a_\mu^{\rm hvp} = \int_0^\infty ds \, w(s) \, R(s)$ with $w(m_\rho^2) = 1.624\times 10^{-8} {\rm GeV}^{-2}$,
one obtains the fairly realistic number
\be
a_\mu^{\rm hvp}(\rho) = 468 \times 10^{-10}.
\ee

In the following, we consider the implications of one-loop corrections.

\subsection{The photonless Lagrangian including explicit counterterms}

We work in the theory without a photon ($A_\mu=0$) and renormalize it at one-loop order.
The Lagrangian is then
\ba
{\cal L}_E &=&   \frac{1}{4} F_{\mu\nu}(\rho)^2 + \frac{1}{2} m_\rho^2 \rho_\mu^2
 + (D_\mu\pi)^\dagger (D_\mu\pi) + m_\pi^2 \pi^\dagger \pi
\nonumber \\ && + \frac{1}{4}(Z_3-1) F_{\mu\nu}^2 + \frac{1}{2}\delta m_\rho^2 \rho_\mu^2
 + (Z_2-1) (\partial_\mu \pi^\dagger \partial_\mu \pi + m_\pi^2 \pi^\dagger \pi) + Z_2 \delta m_\pi^2 \pi^\dagger \pi
\nonumber \\ &&  + g(Z_1-1) \rho_\mu j_\mu.
\ea
The pion-loop contribution $\Pi(k^2)$
appears in the two-point function of the $\rho_\mu$ field. To one-loop order, ignoring the counterterms for now,
\ba
\<\rho_\mu(x) \rho_\nu(y)\> &=& G_{\mu\nu}(x) 
 + \frac{g^2}{2} \<\rho_\mu(x) \int_z \rho_\lambda(z) j_\lambda(z) \int_w \rho_\sigma(w) j_\sigma(w) \rho_\nu(y)\>
\\ && - g^2 \<\rho_\mu(x) \int_z \rho_\lambda(z) \rho_\lambda(z) \pi^\dagger(z) \pi(z)\;\rho_\nu(y)\>
\nonumber\\ &=& \int \frac{d^dk}{(2\pi)^d} \,e^{ik(x-y)}\Big\{\frac{\delta_{\mu\nu} + k_\mu k_\nu /m_\rho^2}{k^2 + m_\rho^2}
+ \frac{\delta_{\mu\lambda} + k_\mu k_\lambda /m_\rho^2}{k^2 + m_\rho^2} \;
g^2 \Pi(k^2) \frac{k^2 \delta_{\lambda\nu} - k_\lambda k_\nu}{k^2 + m_\rho^2}\Big\}.
\nonumber
\ea
Performing the resummation of the geometric series,
\ba
\<\rho_\mu(x) \rho_\nu(y)\> &=& \int \frac{d^dk}{(2\pi)^d} e^{ik(x-y)}\Big\{\frac{\delta_{\mu\nu} + k_\mu k_\nu /m_\rho^2}{k^2 + m_\rho^2}
+ \frac{\delta_{\mu\lambda} + k_\mu k_\lambda /m_\rho^2}{k^2 + m_\rho^2} \;
g^2 \Pi(k^2) \frac{k^2 \delta_{\lambda\nu} - k_\lambda k_\nu}{k^2 + m_\rho^2}
\nonumber\\ && + \frac{\delta_{\mu\lambda} + k_\mu k_\lambda /m_\rho^2}{k^2 + m_\rho^2} \;
\Big(g^2 \Pi(k^2) \frac{k^2 \delta_{\lambda\nu} - k_\lambda k_\nu}{k^2 + m_\rho^2}\Big)^2 + \dots\Big\}
\nonumber \\ &=& \int \frac{d^dk}{(2\pi)^d} \,e^{ik(x-y)}\, \left(M(k)  (1-T(k))^{-1} \right)_{\mu\nu},
\ea
where
\ba
M_{\mu\nu}(k) &=& \frac{\delta_{\mu\nu} + k_\mu k_\nu /m_\rho^2}{k^2 + m_\rho^2},
\\
T_{\lambda\nu}(k) &=& g^2 \Pi(k^2) \frac{k^2 \delta_{\lambda\nu} - k_\lambda k_\nu}{k^2 + m_\rho^2}.
\ea
Since $1-T$ is of the form
\be
1-T = 1-f + f\hat k \hat k^\top, \qquad f = g^2 \Pi(k^2)\frac{k^2}{k^2 + m_\rho^2}
\ee
($\hat k = k/|k|$), its inverse is given by
\be
(1-T)^{-1} = 1 - f \hat	k \hat k^\top.
\ee
Thus one finds
\be
\left(M(k)  (1-T(k))^{-1} \right)_{\mu\nu}
= \frac{\delta_{\mu\nu} + k_\mu k_\nu (1-g^2 \Pi(k^2))/ m_\rho^2 }{k^2(1- g^2 \Pi(k^2))+m_\rho^2}.
\ee
Taking into account the counterterms, one finds
\be
\int d^dx \;e^{ik(x-y)}\,\<\rho_\mu(x) \rho_\nu(y)\> =
\frac{\delta_{\mu\nu} + k_\mu k_\nu (Z_3-g^2 \Pi(k^2))/ (m_\rho^2+\delta m_\rho^2) }{k^2(Z_3 - g^2 \Pi(k^2))+m_\rho^2+ \delta m_\rho^2}.
\ee
The renormalization conditions we impose on the denominator are,
\ba
\Big(k^2(Z_3 - g^2 {\rm Re}\,\Pi(k^2))+m_\rho^2+ \delta m_\rho^2 \Big)_{k^2=-m_\rho^2} &=& 0,
\\
\frac{d}{dk^2} \Big(k^2(Z_3 - g^2 {\rm Re}\,\Pi(k^2))+m_\rho^2+ \delta m_\rho^2 \Big|_{k^2=-m_\rho^2} \Big)_{k^2=-m_\rho^2} &=& 1.
\ea
These conditions lead to the finite $\rho$ mass shift
\ba
\delta m_\rho^2 = m_\rho^2 \;g^2\,\Big(k^2 \frac{d}{dk^2}{\rm Re}\,\Pi(k^2)\Big)_{k^2=-m_\rho^2},
\ea
and the wave-function renormalization
\be
Z_3-1 =  g^2 \Big(1 + k^2 \frac{d}{dk^2}\Big)\,{\rm Re}\,\Pi(k^2)\Big|_{k^2=-m_\rho^2} .
\ee

\subsection{Counterterm for the kinetic mixing term}

In addition to the counterterms treated above, the $g_\gamma^{-1}$ coupling gets renormalized by the pion-loop.
Thus we must add the counterterm
\be\label{eq:kinmixct}
\delta {\cal L}_E = \frac{e}{2} \delta\frac{1}{g_\gamma} F_{\mu\nu}(A) F_{\mu\nu}(\rho)
\ee
to the Lagrangian 
of Eq.\ (\ref{eq:LEinit}). Starting from that Lagrangian, one obtains to one-loop order
\ba
\<A_\mu(x) \,\rho_\nu(y)\> &=& \frac{-e}{2} \<A_\mu(x) \left(\frac{1}{g_\gamma} +\delta\frac{1}{g_\gamma}\right)
\int_z F_{\lambda\sigma}(A)_z F_{\lambda\sigma}(\rho)_z\,\rho_\nu(y)\>_0
\\ \nonumber&& + \<A_\mu(x) e\int_z A_\lambda(z) j_\lambda(z) \;g\int_w \rho_\sigma(w) j_\sigma(w)\,\rho_\nu(y)\>
\\ \nonumber&& + \<A_\mu(x) (-2eg)\int_z A_\lambda(z) \rho_\lambda(z) \pi^*(z)\pi(z)\,\rho_\nu(y)\>
\\ \nonumber&=& e \int\frac{d^dk}{(2\pi)^d}\frac{e^{ik(x-y)}}{k^2(k^2+m_\rho^2)}(\delta_{\mu\nu} k^2 -k_\mu k_\nu)
\Big( - \frac{1}{g_\gamma} - \delta \frac{1}{g_\gamma} + g \Pi(k^2)\Big).
\ea
Here, we require that at $k^2=-m_\rho^2$, the correlation function be given by its tree-level value,
and therefore set
\be
\delta \frac{1}{g_\gamma}  = g \;{\rm Re}\,\Pi(-m_\rho^2).
\ee

Having determined all the required counterterms at one-loop order,
we consider the quantity that is computed in lattice~QCD, namely the photon two-point function,
evaluated at $A_\mu=0$.

\subsection{The current-current correlator}

The current-current correlator computed in lattice~QCD corresponds to $\frac{\delta^2 \log Z[A]}{\partial A_\mu(x) \partial A_\nu(y)}$,
and this must be matched to the Sakurai QFT. In this context we regard $A_\mu(x)$ as a background field for the remaining
degrees of freedom, $\pi$ and $\tilde \rho_\mu$. We note
\be
\frac{\delta S}{\delta A_\mu(x)} =
\frac{e}{g_\gamma} \partial_\alpha F_{\mu\alpha}(\rho)
-ie \left(\pi \partial_\mu \pi^* - \pi^* \partial_\mu\pi\right)
 + 2e g \rho_\mu \pi^*\pi + 2e^2 A_\mu \pi^*\pi.
 \ee
Let
 \be
j_\mu = -i \left(\pi \partial_\mu \pi^* - \pi^* \partial_\mu\pi\right)
\ee
be the electromagnetic current carried by the charged pions.
Then
\ba
\left.\frac{\delta^2 \log Z[A]}{\partial A_\mu(x) \partial A_\nu(y)}\right|_{A=0}
&=& \left.
\Big\< \frac{\delta S}{\delta A_\mu(x)} \frac{\delta S}{\delta A_\nu(y)} \Big\>_{\rm conn}
- \Big\< \frac{\delta^2 S}{\partial A_\mu(x) \partial A_\nu(y)} \Big\>\right|_{A=0}
\\ &=&  
e^2 \Big\< \Big(j_\mu + \frac{1}{g_\gamma}\partial_\alpha F_{\mu\alpha}(\rho) + 2 g \rho_\mu \pi^*\pi \Big)_x
\Big(j_\nu + \frac{1}{g_\gamma}\partial_\beta F_{\nu\beta}(\rho)+ 2 g \rho_\nu \pi^*\pi\Big)_y \Big\>
\nonumber \\ && - 2e^2 \delta_{\mu\nu} \delta(x-y) \<\pi^*\pi\>,
\ea
where the expectation value is now in the theory without the field $A_\mu$.
Evaluating the expectation value to order $g^0$, counting $g_\gamma/g$ to be O(1), yields
\ba \la{eq:fin_expr}
\frac{1}{e^2}\left.\frac{\delta^2 \log Z[A]}{\partial A_\mu(x) \partial A_\nu(y)}\right|_{A=0}
&=& 
 \frac{1}{g_\gamma^2}  \<\partial_\alpha F_{\mu\alpha}(\rho)_x \partial_\beta F_{\nu\beta}(\rho)_y \>_0
 \\ && + \<j_\mu(x) j_\nu(y)\>_{\rm sQED} - 2 \delta_{\mu\nu} \delta(x-y) G_{m_\pi}(0)
\nonumber \\ && + 2\frac{g}{g_\gamma} G_{m_\pi}(0) \Big( \< \partial_\alpha F_{\mu\alpha}(\rho)_x \rho_\nu(y)\>_0
+ \< \rho_\mu(x) \partial_\beta F_{\nu\beta}(\rho)_y \>_0 \Big)
\nonumber \\ && +  \frac{1}{2}\frac{g^2}{g_\gamma^2}  \Big\<\partial_\alpha F_{\mu\alpha}(\rho)_x\int_z \rho_\sigma(z)j_\sigma(z)
\int_w \rho_\lambda(w)j_\lambda(w)\partial_\beta F_{\nu\beta}(\rho)_y\Big\>_0
\nonumber \\ && - \frac{g^2}{g_\gamma^2}\Big\<\partial_\alpha F_{\mu\alpha}(\rho)_x \int_z \rho_\sigma(z)\rho_\sigma(z) \pi^*(z)\pi(z)
\partial_\beta F_{\nu\beta}(\rho)\Big\>_0
\nonumber \\ && - \frac{g}{g_\gamma} \Big\< \partial_\alpha F_{\mu\alpha}(\rho)_x \int_z \rho_\sigma(z)j_\sigma(z)  j_\nu(y)
+ j_\mu(x) \int_z \rho_\sigma(z)j_\sigma(z)  \partial_\beta F_{\nu\beta}(\rho)_y\Big\>_0.
\nonumber
\ea

For the first line of Eq.\ (\ref{eq:fin_expr}), one derives from the massive vector propagator expression (\ref{eq:Gv})
\ba
\<\partial_\alpha F_{\mu\alpha}(x) \rho_\nu(0)\> = \delta_{\mu\nu} \delta(x) - m_\rho^2 G_{\mu\nu}(x)
\ea
and
\ba
\<\partial_\alpha F_{\mu\alpha}(x) \partial_\beta F_{\nu\beta}(0)\> = m_\rho^4 G_{\mu\nu}(x) +
\Big(\partial_\mu \partial_\nu - (\triangle + m_\rho^2) \delta_{\mu\nu}\Big) \delta(x).
\ea
Secondly, the scalar QED contribution reads
\be
\<j_\mu(x) j_\nu(0)\>_{\rm sQED} = 2\Big(\partial_\mu G_{m_\pi}(x) \partial_\nu G_{m_\pi}(x) -  G_{m_\pi}(x) \partial_\mu\partial_\nu  G_{m_\pi}(x)\Big).
\ee
Now to the one-loop contribution of the last line of Eq.\ (\ref{eq:fin_expr}),
along with the corresponding tadpole contribution of the third line.
Noting that
\be
\Pi^{\rm sQED}_{\mu\nu}(k)\equiv \int d^dx \,e^{ikx} \, \Big(\<j_\mu(x)j_\nu(0)\>_{\rm sQED} - 2 \delta_{\mu\nu} \delta(x) G_{m_\pi}(0)\Big)
= (\delta_{\mu\nu}k^2 - k_\mu k_\nu) \Pi(k^2)
\ee
with 
\be
\Pi(k^2) = - \frac{1}{(4\pi)^{d/2}} \Gamma(2-{\txts\frac{d}{2}}) \int_0^1 dx\frac{(1-2x)^2}{(x(1-x)k^2+m^2)^{2-d/2}},
\ee
we obtain
\ba\la{eq:3rd.last}
&& + 2\frac{g}{g_\gamma} G_{m_\pi}(0)  \< \partial_\alpha F_{\mu\alpha}(\rho)_x \rho_\nu(y)\>_0
- \frac{g}{g_\gamma} \Big\< \partial_\alpha F_{\mu\alpha}(\rho)_x \int_z \rho_\sigma(z)j_\sigma(z)  j_\nu(y)\Big\>_0
\\ && = 2\frac{g}{g_\gamma} G_{m_\pi}(0) \delta_{\mu\nu} \delta(x-y)
- \frac{g}{g_\gamma} \<j_\mu(x)j_\nu(y)\>_{\rm sQED}
\nonumber\\ && + m_\rho^2 \frac{g}{g_\gamma} \int \frac{d^dk}{(2\pi)^d} \frac{e^{ik(x-y)}}{k^2+m_\rho^2} (\delta_{\mu\nu}k^2 - k_\mu k_\nu) \Pi(k^2),
\nonumber
\ea
The other terms on line three and on the last line of Eq.\ (\ref{eq:fin_expr})
simply correspond to the exchange $(x,\mu) \leftrightarrow(y,\nu)$,  thus simply leading to
doubling the contribution of Eq.\ (\ref{eq:3rd.last}).

Lines four and five of Eq.\ (\ref{eq:fin_expr}) are best handled together and we find
\ba
 &&   \frac{1}{2}\frac{g^2}{g_\gamma^2}  \Big\<\partial_\alpha F_{\mu\alpha}(\rho)_x\int_z \rho_\sigma(z)j_\sigma(z)
\int_w \rho_\lambda(w)j_\lambda(w)\partial_\beta F_{\nu\beta}(\rho)_y\Big\>_0
\nonumber \\ && - \frac{g^2}{g_\gamma^2}\Big\<\partial_\alpha F_{\mu\alpha}(\rho)_x \int_z \rho_\sigma(z)\rho_\sigma(z) \pi^*(z)\pi(z)
\partial_\beta F_{\nu\beta}(\rho)\Big\>_0
\nonumber\\ && = \frac{g^2}{g_\gamma^2}\Big\{ \<j_\mu(x) j_\nu(y)\>_{\rm sQED} -2 G_{m_\pi}(0) \delta_{\mu\nu} \delta(x-y)
 \nonumber\\ && -2m_\rho^2 \int \frac{d^dk}{(2\pi)^d} \frac{e^{ik(x-y)}}{k^2+m_\rho^2} (\delta_{\mu\nu}k^2 - k_\mu k_\nu) \,\Pi(k^2)
\nonumber\\ && + m_\rho^4 \int \frac{d^dk}{(2\pi)^d} \frac{e^{ik(x-y)}}{(k^2+m_\rho^2)^2} (\delta_{\mu\nu}k^2 - k_\mu k_\nu) \,\Pi(k^2) \Big\}.
\ea
Altogether, we have
\ba \la{eq:fin_expr2}
\frac{1}{e^2}\left.\frac{\delta^2 \log Z[A]}{\partial A_\mu(x) \partial A_\nu(y)}\right|_{A=0}
&=&  \frac{m_\rho^4}{g_\gamma^2} G_{\mu\nu}(x-y) 
  + \frac{1}{g_\gamma^2} \Big(\partial_\mu \partial_\nu - (\triangle + m_\rho^2) \delta_{\mu\nu}\Big) \delta(x-y)
\nonumber\\ && +  \Big(1-\frac{g}{g_\gamma} \Big)^2 \Big(\<j_\mu(x) j_\nu(y)\>_{\rm sQED} - 2 \delta_{\mu\nu} \delta(x-y) G_{m_\pi}(0)\Big)
\nonumber\\ && + 2m_\rho^2 \frac{g}{g_\gamma} \Big(1-\frac{g}{g_\gamma}\Big) ( \partial_\mu \partial_\nu  -\delta_{\mu\nu}\triangle )
\int \frac{d^dk}{(2\pi)^d} \frac{e^{ik(x-y)}}{k^2+m_\rho^2}  \Pi(k^2)
\nonumber \\ && + \frac{g^2}{g_\gamma^2}  (\partial_\mu \partial_\nu  -\delta_{\mu\nu}\triangle)
m_\rho^4  \int \frac{d^dk}{(2\pi)^d} \frac{e^{ik(x-y)}}{(k^2+m_\rho^2)^2} \, \Pi(k^2)\,.
\la{eq:fin_expr3}
\ea
Each term is transverse, i.e.\ yields zero when $\partial_\mu^{(x)}$ is applied to it;
for the first term, note that
\be
\partial_\nu G_{\mu\nu}(x) = \frac{1}{m_\rho^2} \,\partial_\mu \delta(x).
\ee

\subsubsection{Contribution of counterterms to the current-current correlator}

The contribution of the counterterm (\ref{eq:kinmixct}) amounts to replacing
$g_\gamma^{-1}$ by $(g_\gamma^{-1}+\delta g_\gamma^{-1})$.  Since the
counterterm represents a relative correction of order $g^2$, this
correction needs be applied only to the leading terms, namely those of
order $g_\gamma^{-2}$. Thus we obtain the contribution
\be
\frac{1}{e^2}\left.\frac{\delta^2 \log Z[A]}{\partial A_\mu(x) \partial A_\nu(y)}\right|_{A=0}
= \dots + \frac{2}{g_\gamma}\delta \frac{1}{g_\gamma} m_\rho^4 G_{\mu\nu}(x-y)
+ \frac{2}{g_\gamma} \delta \frac{1}{g_\gamma}  \Big(\partial_\mu \partial_\nu - (\triangle + m_\rho^2) \delta_{\mu\nu}\Big) \delta(x-y)
+ \dots
\ee

The contribution of the counterterms proportional to $(Z_3-1)$ and $\delta m_\rho^2$ reads
\ba
\frac{1}{g_\gamma^2} \< \partial_\alpha F_{\mu\alpha}(x) \,\partial_\beta F_{\nu\beta}(y)\>_{\rm c.t.}
&=& -\frac{1}{g_\gamma^2} \<	\partial_\alpha	F_{\mu\alpha}(x) \,\int_z \frac{\delta m_\rho^2}{2} \rho_\lambda(z)\rho_\lambda(z)\,
 \partial_\beta F_{\nu\beta}(y)\>_0
\nonumber \\ && -\frac{1}{g_\gamma^2} \< \partial_\alpha F_{\mu\alpha}(x)\,(Z_3-1)\,\int_z \frac{1}{4} F_{\lambda\sigma}(z)F_{\lambda\sigma}(z)
\; \partial_\beta F_{\nu\beta}(y)\>_0
\nonumber \\ &=& - \frac{\delta m_\rho^2}{g_\gamma^2} \Big( \delta_{\mu\nu} \delta(x-y) - 2m_\rho^2 G_{\mu\nu}(x-y)
\nonumber\\ && \qquad  +  \int_k \frac{e^{ik(x-y)}}{(k^2+m_\rho^2)^2} (\delta_{\mu\nu} m_\rho^4 +  k_\mu k_\nu(2m_\rho^2 + k^2)) \Big)
\nonumber\\ && - \frac{Z_3-1}{g_\gamma^2} \Big( \big(\delta_{\mu\nu}(-\triangle^{(x)} - 2m_\rho^2) + \partial_\mu^{(x)}\partial_\nu^{(x)}\big) \delta(x-y)
\nonumber \\ && + 2m_\rho^4 G_{\mu\nu}(x-y)
 + m_\rho^4 \int_k \frac{e^{ik(x-y)}}{(k^2+m_\rho^2)^2} (k^2 \delta_{\mu\nu} - k_\mu k_\nu)\Big)\,.
 \qquad 
\ea
Thus we have the final result, now including all counterterm contributions,
\ba
\frac{1}{e^2}\left.\frac{\delta^2 \log Z[A]}{\partial A_\mu(x) \partial A_\nu(y)}\right|_{A=0}
&=&  \frac{m_\rho^2(m_\rho^2+2\delta m_\rho^2)}{g_\gamma^2} G_{\mu\nu}(x-y) 
  + \frac{1}{g_\gamma^2} \Big(\partial_\mu \partial_\nu - (\triangle + m_\rho^2) \delta_{\mu\nu}\Big) \delta(x-y)
\nonumber\\ && +  \Big(1-\frac{g}{g_\gamma} \Big)^2 \Big(\<j_\mu(x) j_\nu(y)\>_{\rm sQED} - 2 \delta_{\mu\nu} \delta(x-y) G_{m_\pi}(0)\Big)
\nonumber\\ && + 2m_\rho^2   ( \partial_\mu \partial_\nu  -\delta_{\mu\nu}\triangle )
\int \frac{d^dk}{(2\pi)^d} \frac{e^{ik(x-y)}}{k^2+m_\rho^2}
\Big( \frac{g}{g_\gamma}(1-\frac{g}{g_\gamma}) \Pi(k^2) - \frac{\delta g_\gamma^{-1}}{g_\gamma}  + \frac{Z_3-1}{g_\gamma^2} \Big)
\nonumber \\ && + \frac{1}{g_\gamma^2}  (\partial_\mu \partial_\nu  -\delta_{\mu\nu}\triangle)
m_\rho^4  \int \frac{d^dk}{(2\pi)^d} \frac{e^{ik(x-y)}}{(k^2+m_\rho^2)^2} \, (g^2 \Pi(k^2) - (Z_3-1) )
\nonumber\\ && - \frac{\delta m_\rho^2}{g_\gamma^2} \Big( \delta_{\mu\nu} \delta(x-y) 
 +  \int_k \frac{e^{ik(x-y)}}{(k^2+m_\rho^2)^2} (\delta_{\mu\nu} m_\rho^4 +  k_\mu k_\nu(2m_\rho^2 + k^2)) \Big)
  \nonumber \\ &&
+ \Big(\frac{2}{g_\gamma} \delta \frac{1}{g_\gamma} - \frac{Z_3-1}{g_\gamma^2}\Big)  \Big(\partial_\mu \partial_\nu - \triangle  \delta_{\mu\nu}\Big) \delta(x-y)\,.
\la{eq:fin_expr4}
\ea


\subsection{Finite-size effects on the current-current correlator}

Let
\be
C_{\mu\nu}(x) = -\< j_\mu(x) j_\nu(0)\>
\ee
be the Euclidean position-space vector correlator.

Consider the case of $g=g_\gamma$. Then the only term contributing to finite-size effects
which is not suppressed by $e^{-m_V L/2}$ is
\be
C^{(L)}_{\mu\nu}(x) = \frac{m_\rho^4}{V}\sum_k e^{ikx} \,\frac{\overline\Pi^{(L)}_{\mu\nu}(k)}{(k^2+m_\rho^2)^2},
\ee
where $\overline\Pi^{(L)}_{\mu\nu}(k)$ is the finite-volume renormalized vacuum polarization tensor; note that the renormalization
is always performed in infinite volume.
It is useful to decompose the latter into its infinite-volume counterpart, plus a remainder,
\be
\overline\Pi^{(L)}_{\mu\nu}(k) = (\delta_{\mu\nu} k^2 - k_\mu k_\nu) (\Pi(k^2)- (Z_3-1)/g^2) + \Delta\Pi^{(L)}_{\mu\nu}(k),
\ee
because the remainder is ultraviolet finite.
We can then write
\ba
\label{fse_sakurai_winding_formula}
C^{(L)}_{\mu\nu}(x) - C^{(\infty)}_{\mu\nu}(x) &=&
\sum_{n\in \mathbb{Z}^4\backslash\{ 0\}}  C^{(\infty)}_{\mu\nu}(x+nL)
\nonumber\\ &&+ \sum_{n\in \mathbb{Z}^4} \int  \frac{d^4k}{(2\pi)^4} e^{ik(x+nL)} \frac{m_\rho^4}{(k^2+m_\rho^2)^2}
\Delta\Pi^{(L)}_{\mu\nu}(k).
\ea

It is instructive and useful to compute $C_{\mu\nu}(x)$ in infinite volume,
\be
C^{(\infty)}_{\mu\nu}(x) = (\partial_\mu \partial_\nu - \delta_{\mu\nu} \triangle)
\int \frac{d^4k}{(2\pi)^4}\, e^{ikx} \, \frac{m_\rho^4}{(k^2+m_\rho^2)^2}\, \left(\Pi(k^2)-\Pi(0) + \left(\Pi(0)- \frac{Z_3-1}{g^2}\right)\right) .
\ee
It is clear that the $(\Pi(0)- \frac{Z_3-1}{g^2})$ term is rapidly decaying in position space, being O($e^{-m_V |x|}$).
To calculate the position-space dependence of the other term, insert the spectral representation
\be
\Pi(k^2)-\Pi(0) = k^2 \int_0^\infty ds \;\frac{\rho_{\rm sQED}(s)}{s(s+k^2)} ,
\ee
with the spectral function normalized according to
\be
\rho(s) = R(s)/(12\pi^2)\stackrel{\rm sQED}{=} \frac{1}{48\pi^2} (1-4m_\pi^2/s)^{3/2}.
\ee
In this form, the $d^4k$ integral can be performed,
\ba
C^{\infty}_{\mu\nu}(x) &=& m_\rho^4 (\partial_\mu \partial_\nu - \delta_{\mu\nu} \triangle)
\Big\{ - (\Pi(0)- \frac{Z_3-1}{g^2}) \frac{\partial}{\partial m_\rho^2} G_{m_\rho}(x) 
 \\ && +\int_0^\infty \frac{ds}{s(s-m_\rho^2)}\,\rho_{\rm sQED}(s) \Big( \frac{-s}{s-m_\rho^2} (G_{\sqrt{s}}(x) - G_{m_\rho}(x))
 + m_\rho^2 \frac{\partial}{\partial m_\rho^2} G_{m_\rho}(x) \Big)\Big\}.
 \nonumber
\ea

For the second term, we note that
\be
\Delta\Pi^{(L)}_{\mu\nu}(k) = \sum_{\nu\neq 0} \int \frac{d^4q}{(2\pi)^4}\, e^{iLq\cdot \nu}
\frac{-(k+2q)_\mu (k+2q)_\nu + 2\delta_{\mu\nu} ((k+q)^2 + m_\pi^2)}{((k+q)^2+m_\pi^2) (q^2+m_\pi^2)}.
\ee
One finds, using a Feynman parameter $\alpha$,
\ba
\Delta\Pi^{(L)}_{\mu\nu}(k) &=& \sum_{\nu\neq 0} \Delta\Pi^{(L)}_{\mu\nu}(k,y=L\nu),
\\
\Delta\Pi^{(L)}_{\mu\nu}(k,y) &=& \{-(k+2q)_\mu (k+2q)_\nu + 2\delta_{\mu\nu} ((k+q)^2 + m_\pi^2)\}_{q=-i\nabla_y}
\nonumber \\ && \cdot \frac{1}{8\pi^2}\int_0^1 d\alpha \,e^{-i\alpha k\cdot y}\, K_0(\sqrt{\alpha(1-\alpha)k^2+m_\pi^2}|y|).
\ea
For the next step, the $d^4k$ integral can be reduced to a one-dimensional integral,
\ba
&& \int \frac{d^4k}{(2\pi)^4}\,e^{ikx} \frac{m_\rho^4}{(k^2+m_\rho^2)^2}\;\Delta\Pi^{(L)}_{\mu\nu}(k,y)
= \frac{m_\rho^4}{32\pi^4}
\nonumber  \\ && \cdot \{-(k+2q)_\mu (k+2q)_\nu + 2\delta_{\mu\nu} ((k+q)^2 + m_\pi^2)\}_{q=-i\nabla_y,k=-i\nabla_x}
\nonumber  \\ && \cdot \int_0^1 \frac{d\alpha }{|x-\alpha y|} \int_0^\infty dk \; \frac{k^2}{(k^2+m_\rho^2)^2} \,
J_1(k|x-\alpha y|)\,K_0(\sqrt{\alpha(1-\alpha)k^2+m_\pi^2}|y|).
\ea
Without the $(k^2+m_\rho^2)^{-2}$ factor, the $k$ integral could be performed,
\ba
&& \frac{1}{|x-\alpha y|} \int_0^\infty dk \; k^2 \,
J_1(k|x-\alpha y|)\,K_0(\sqrt{\alpha(1-\alpha)k^2+m_\pi^2}|y|)
\\ && = \frac{m_\pi^2}{\alpha(1-\alpha)} \,\frac{ K_2(\frac{m_\pi}{\sqrt{\alpha(1-\alpha)}}\sqrt{\alpha(y^2-2x\cdot y)+x^2})}{\alpha(y^2-2x\cdot y)+x^2}.
\nonumber
\ea 

Using this result, the actually required integral can be brought into the following, non-oscillatory form,
\ba
&& \frac{1 }{|x-\alpha y|} \int_0^\infty dk \; \frac{k^2}{(k^2+m_\rho^2)^2} \,
J_1(k|x-\alpha y|)\,K_0(\sqrt{\alpha(1-\alpha)k^2+m_\pi^2}|y|)
\\ && = -\frac{\partial}{\partial m_\rho^2} 2\pi^2 \frac{m_\pi^2}{\alpha(1-\alpha)} \int_0^\infty dz\,z^3 \gamma_0^{(m_\rho)}(|x-\alpha y|,z)
\frac{ K_2(\frac{m_\pi}{\sqrt{\alpha(1-\alpha)}}\sqrt{z^2 + \alpha(1-\alpha) y^2})}{z^2 + \alpha(1-\alpha) y^2},
\nonumber
\ea
where
\be
\gamma_n^{(m)}(|x|,|u|) = \frac{n + 1}{2\pi^2|x||u|} \Big(\theta(|x|-|u|) I_{n+1}(m|u|) K_{n+1}(m|x|)
+ \theta(|u|-|x|) I_{n+1}(m|x|) K_{n+1}(m|u|)\Big)
\ee
is the $n^{\rm th}$ coefficient in the Gegenbauer polynomial expansion of the scalar propagator in four dimensions,
\be
G_m(x-u) = \sum_{n\geq 0} \gamma^{(m)}_n(|x|,|u|)\, C^{(1)}_n(\hat u\cdot \hat x),
\ee
with $C^{(1)}_0(z)=1$, $C^{(1)}_1(z)=2z$, $C^{(1)}_2(z)=4z^2-1$  etc.


%% file: tables.tex
\newpage
\section{Tables}

This section provides tables for the  finite-size corrections applied ensemble-by-ensemble (Tab.~\ref{table:fs_correction}), as well as for the correction to reach the reference point in the $(m_\pi,m_K)$ plane (Tab.~\ref{table:corrections_to_ref}).

\begin{table}[h!]
\centering
\begin{tabular}{|cc|cccccc|c|}
\toprule
\multicolumn{1}{|l}{} & \multicolumn{1}{l|}{} & \multicolumn{6}{c|}{Wrap-around-the-world correction}                                                                                                                      & Trunc. correction           \\ \hline
                      &                       &               & Sakurai QFT                 & \multicolumn{1}{c|}{}                            &               & Scalar QED                  &                             & Sakurai QFT                 \\
Id                    & $L$ {[}fm{]}          & $H_{\mu \nu}$ & $H_{\mu \nu}^{\textrm{TL}}$ & \multicolumn{1}{c|}{$H_{\mu \nu}^{\textrm{XX}}$} & $H_{\mu \nu}$ & $H_{\mu \nu}^{\textrm{TL}}$ & $H_{\mu \nu}^{\textrm{XX}}$ & $H_{\mu \nu}^{\textrm{TL}}$ \\ \hline
U102                  & 2.1                   & -3.859(455)   & -2.834 (962)                & \multicolumn{1}{c|}{-5.805(1865)}                & 1.023(320)    & -6.62(155)                  & -10.099(177)                & -1.511                      \\ \hline
H102                  & 2.8                   & -0.990(103)   & -0.759(208)                 & \multicolumn{1}{c|}{-1.243(308)}                 & 0.419(9)      & -1.424(10)                  & -1.577(8)                   & -0.584                      \\ \hline
S400                  & 2.4                   & -2.047(255)   & -1.479(429)                 & \multicolumn{1}{c|}{-2.677(719)}                 & 0.705(28)     & -3.075(38)                  & -3.897(35)                  & -1.654                      \\ \hline
N203                  & 3.1                   & 0.114(26)    & -0.200(21)                  & \multicolumn{1}{c|}{-0.259(49)}                  & 0.266(30)     & -0.747(40)                  & -0.744(31)                  & -0.200                      \\ \hline
N302                  & 2.4                   & -2.396(310)   & -1.693(533)                 & \multicolumn{1}{c|}{-3.130(894)}                 & 0.714(163)    & -3.611(51)                  & -4.712(48)                  & -1.104                      \\ 
\botrule
\end{tabular}
\caption{Results for the corrections for the finite-size effect of discretized momenta calculated in the Sakurai QFT and scalar QED and truncation of the integrand calculated in the Sakurai QFT. 
	See the text in Sect.~\ref{sect:fsescheme} for details. All values are in units of $10^{-10}$.
	To get the FSE correction for N302 with the TL-kernel for example, we have $\textrm{FSE}=-1.693 - 1.104 =-2.797$ for the central value with an uncertainty of $\sigma_{\rm{FSE}}=\sqrt{0.533^2 +
	(0.25\cdot 2.797)^2}= 0.879$}
\label{table:fs_correction}
\end{table}

\begin{table}[h!]
\begin{tabular}{|c|cc|cc|}
\toprule
& \multicolumn{2}{c|}{isovector} & \multicolumn{2}{c|}{strange} \\
Id                                                                  & (LL)          & (CL)          & (LL)         & (CL)         \\ \hline
H102                                                                & 0.11(4)       & 0.12(4)       & -0.5(1)      & -0.49(1)     \\ \hline
S400                                                                & -0.07(2)      & -0.06(2)      & -0.27(1)     & -0.26(1)     \\ \hline
N203                                                                & -0.74(5)      & -0.72(5)      & -0.21(1)     & -0.2(1)      \\ \hline
N302                                                                & -0.63(1)      & -0.62(1)      & 0.22(0)      & 0.22(0)      \\ \hline
\botrule
\end{tabular}
\caption{Corrections to the reference point $m_\pi=350$ MeV $m_K = 450$ MeV determined using calculations based on the TMR method. All values are in units of $10^{-10}$}
\label{table:corrections_to_ref}
\end{table}